\documentclass[10pt,aps,amsfonts,amssymb,longbibliography,twocolumn]{revtex4-1}
\usepackage{mathrsfs} 
\usepackage {amsmath}
\usepackage{graphicx,color}
\usepackage{bm}
\begin{document}

\title{Skyrmion crystal in the RKKY system on the two-dimensional triangular lattice}

\author{Kota Mitsumoto}
\author{Hikaru Kawamura}

\affiliation{Molecular Photoscience Research Center, Kobe University, Kobe 657-8501, Japan}

\begin{abstract}
We study the ordering properties of the isotropic RKKY Heisenberg model on the two-dimensional (2D) triangular lattice by extensive Monte Carlo simulations to get insights into the chiral-degenerate skyrmion crystal (SkX) of metallic magnets.
Our Hamiltonian contains only the spin-quadratic RKKY interaction derived from the spherical Fermi surface, containing neither the nesting nor the many-body interaction. 
The SkX phase is stabilized under applied fields where the frustration associated with the oscillating nature of the RKKY interaction and the emergent many-body interactions generated by thermal fluctuations play important roles. 
\textcolor{black}{Replica symmetry breaking, reported in our recent study on the 3D RKKY model [Phys. Rev. B {\bf 104}, 184432 (2021)], turns out to be absent in the present 2D model.} Implications to the SkX formation mechanism are discussed.
\end{abstract}

\maketitle

\section{Introduction}

Topologically protected spin textures in magnets have attracted much recent attention. In particular, skyrmion, a swirling noncoplanar spin texture whose constituent spin directions wrap a sphere in spin space, has been studied quite extensively. Skyrmion is usually stabilized as a periodic array called the skyrmion crystal (SkX). 
The SkX state was first observed in chiral ferromagnets such as MnSi \cite{muhlbauer2009skyrmion,neubauer2009topological}, FeCoSi \cite{munzer2010skyrmion, yu2010real}, and FeGe \cite{yu2011near} under magnetic fields, where the SkX state is induced by the anti-symmetric Dzyaloshinskii-Moriya (DM) interaction. 
A recent study by Okubo, Chung, and Kawamura \cite{okubo2012multiple} of the $J_1$-$J_3$ ($J_2$) Heisenberg model on the two-dimensional (2D) triangular lattice has revealed that the SkX state can also be realized in centrosymmetric magnets without the DM interaction. This model shows the single-$q$ incommensurate spiral phase in zero field and exhibits the triple-$q$ SkX phase under applied fields at finite temperatures. The SkX state is induced by frustration among nearest- and further-neighbor exchange interactions and stabilized by magnetic fields and thermal fluctuations. Nowadays, the centrosymmetric SkX systems have attracted much interest both experimentally \cite{kurumaji2019skyrmion, takahashi2020competing} and theoretically \cite{leonov2015multiply, hayami2016bubble, ozawa2017zero, hayami2017effective, lin2018face, hayami2019effect, wang2020skyrmion, hayami2020degeneracy, hayami2021noncoplanar, hayami2021square, yambe2021skyrmion, wang2021meron}. 

The skyrmion is characterized by the integer topological charge in units of the solid angle $4\pi$, whose sign represents the swirling direction of the skyrmion, which is also represented by the sign of the scalar spin chirality. 
When the DM interaction induces the SkX, the total chirality is determined to be a definite sign depending on materials. In contrast, in the frustration-induced SkX in centrosymmetric magnets, the total chirality can take both positive and negative values, i.e., either SkX or anti-SkX being selected by the spontaneous symmetry breaking. 
We call such a symmetric state a chiral-degenerate SkX state. The chiral-degenerate SkX might exhibit an interesting electromagnetic response, e.g., the topological Hall effect of both signs. Such a chiral-degenerate nature might also give rise to a new interesting phase called the $Z$ phase, a random domain phase consisting of both SkX and anti-SkX.

 Some candidate materials of the frustration-induced SkX are metallic compounds, Gd$_2$PdSi$_3$ \cite{saha1999magnetic, kurumaji2019skyrmion} and EuCuSb \cite{takahashi2020competing}. The primary interaction between localized magnetic moments is the Ruderman-Kittel-Kasuya-Yosida (RKKY) interaction mediated by conduction electrons \cite{ruderman1954indirect, kasuya1956theory, yosida1957magnetic}. When the exchange coupling $J_{sd}$ between the conduction electron and the localized spin is sufficiently weak compared with the Fermi energy $\epsilon_{\rm F}$, the RKKY interaction can be derived via the second-order perturbation with respect to $J_{sd}/\epsilon_{\rm F}$. The RKKY interaction in metals oscillates in sign with the distance $r$, providing frustration just as in the competing $J_1$-$J_3$ ($J_1$-$J_2$) interaction in insulators, while it is the long-range interaction falling as $1/r^3$ in contrast to the short-range interaction. The similarity in the inherent frustration of the interaction suggests that the chiral-degenerate SkX might also be stabilized in RKKY metals.

\textcolor{black}{The theoretical studies on metallic systems, most of which dealt with the $T=0$ case, have suggested that the SkX cannot be realized in centrosymmetric metals by only the isotropic RKKY interaction.} Refs. \cite{ozawa2017zero, hayami2017effective, hayami2020degeneracy} investigated the SkX phase by numerical simulation on the Kondo lattice model and the analysis of the effective classical spin model in $q$-space, and suggested that incommensurate wavenumbers dictated by the Fermi surface, i.e., nesting, and the four-body spin interactions arising from the higher-order perturbation, especially the biquadratic interaction with the positive coefficient, are essential in stabilizing the SkX. 
Ref. \cite{wang2020skyrmion} investigated the RKKY system mediated by 2D electron gas by the $T=0$ variational method, showing that the SkX state needed an easy-axis anisotropy to be stabilized, in addition to the frustration of the RKKY interaction. Meanwhile, these studies mostly dealt with the $T=0$ case, and the studies of the possible SkX state in metals have been rather scarce at finite temperatures. Since the SkX phase in the isotropic short-range system is stabilized by thermal fluctuations, it is important to explore the possible SkX phase of the RKKY system at finite temperatures.

 In the present paper, we investigate by extensive Monte Carlo (MC) simulations the ordering properties of the isotropic classical Heisenberg model on the 2D triangular lattice interacting via the standard RKKY interaction associated with the spherical Fermi surface. Then, our model does not possess the nesting, the four-body (biquadratic) interaction, or the magnetic anisotropy. Our first aim is to examine whether the SkX state is ever possible, even in such a simple model of metallic systems.

 Our very recent study on the 3D RKKY system \cite{mitsumoto2021replica} has indicated that the SkX is stabilized, but the state accompanies quite an exotic property of the replica-symmetry breaking (RSB) familiar in glassy systems \cite{mezard1987spin, mydosh1993spin, kawamura2015spin} but rarely seen in regularly ordered states \cite{mitsumoto2021replica}. Hence, the SkX state in the 3D RKKY system is quite different in nature from the chiral-degenerate SkX phase observed in the 2D short-range model \cite{okubo2012multiple}. The second aim of the present paper is to clarify whether an RSB phenomenon occurs in the 2D RKKY system.

 By extensive MC simulations, we have shown that the 2D RKKY system certainly exhibits a chiral-degenerate triple-$q$ SkX phase under magnetic fields, which, however, does not accompany the RSB. The results demonstrate that neither the nesting, the four-body interaction, nor the magnetic anisotropy is indispensable for the SkX formation. Implications of the results to the SkX formation mechanism will then be discussed.
 
 \textcolor{black}{The rest of the paper is organized as follows.
 In Sec. \ref{model}, we introduce our model and explain the MC simulation method employed.
 We present the results of our MC simulations in Sec. \ref{result}, including the temperature versus magnetic-field phase diagram in Sec. \ref{phase_diagram}, spin structure factors and real-space spin and chirality configurations in Sec. \ref{structure_factor}, temperature dependences of several physical quantities in Sec. \ref{phys_quantities}, and the distribution functions of the total and staggered scalar chiralities in Sec. \ref{absence}.
Summary and discussion of the results are given in Sec. \ref{summary}, including the relationship between this study and previous studies for skyrmions in centrosymmetric metallic magnets.
 We show some of the details of the Ewald sum method in Appendix \ref{ewald_sum}, the data of the temperature dependences of the physical quantities in the single-$q$ phase in Appendix \ref{low_field}, the definition and the temperature dependence of the $Z_3$-symmetry breaking parameter in Appendix \ref{rho3}, the histograms of the $Z_3$-symmetry breaking parameter in each phase in Appendix \ref{histogram}, the derivation of the four-body interaction induced by thermal fluctuations in Appendix \ref{derivation}, and the evaluation of the small parameter $(J_{sd}/\epsilon_{\rm F})^2$ from the experimental data in Appendix \ref{evaluation}.}

 \section{Model and method}\label{model}
We consider the classical Heisenberg model on the 2D triangular lattice interacting with the long-range RKKY interaction.
Our Hamiltonian is given by
\begin{eqnarray}
H&=&-\sum_{i,j}J_{ij}\bm{S}_i \cdot \bm{S}_j - h \sum_{i=1}^N S_i^{z} , 
\label{hamiltonian} \\
J_{ij} &=& -J_0 \left( \frac{\cos(2k_{\rm F} r_{ij})}{r_{ij}^3} - \frac{\sin(2k_{\rm F} r_{ij})}{2k_{\rm F} r_{ij}^4} \right), 
\label{RKKY}
\end{eqnarray}
where $\bm{S}_i=(S_i^x, S_i^y, S_i^z)$ is a three-component unit vector at the site $i$, $J_{ij}$ is the RKKY interaction between the sites $i$ and $j$ with the energy scale $J_0>0$ ($J_0$ is proportional to $\epsilon_{\rm F}(J/\epsilon_{\rm F})^2$), $r_{ij} = |\bm{r}_i - \bm{r}_j|$ is the distance between the sites $i$ and $j$ in units of the lattice constant $a$, $k_{\rm F}$ is the Fermi wavenumber in units of $a^{-1}$, and $h$ is the magnetic-field intensity. The sum is taken over all spin pairs on the 2D triangular lattice, and the total number of lattice sites is $N=L\times L$. The dimensionless temperature and magnetic field are defined by $\tilde{T} = k_{\rm B}T/J_0$ and $\tilde{h} = h/J_0$, and hereafter we denote $\tilde T$ and $\tilde h$ simply as $T$ and $h$. Note that the RKKY interaction of Eq.~(\ref{RKKY}) assumes the spherical Fermi surface of the 3D electron gas.

\begin{figure}[t]
\includegraphics[clip, width=85mm]{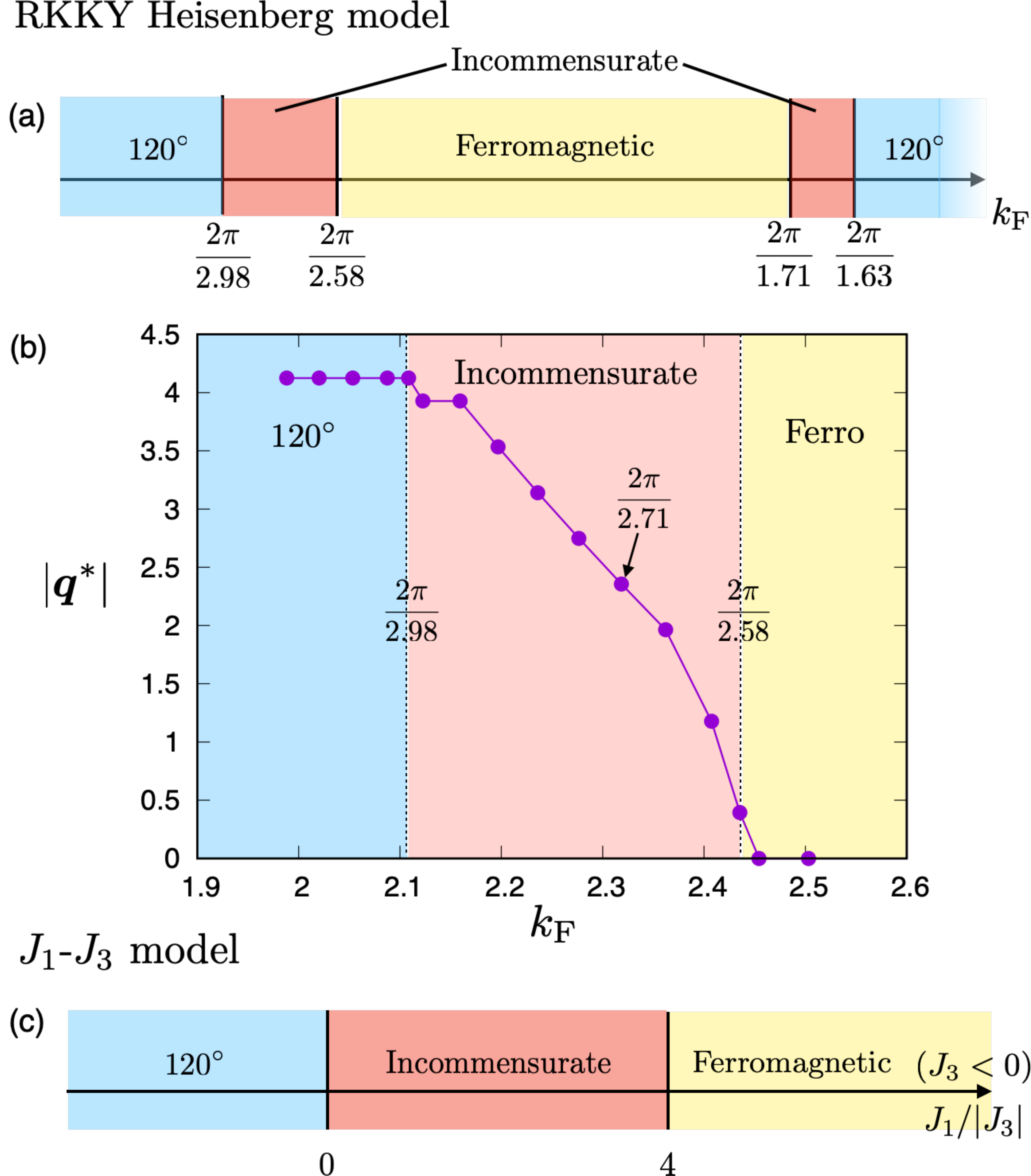}
\caption{The ground-state spin structures of the classical Heisenberg model on the 2D triangular lattice of (a) the RKKY model versus $k_{\rm F}$, and of (c) the $J_1$-$J_3$ model with the ferromagnetic nearest-neighbor coupling $J_1>0$ and the antiferromagnetic third-neighbor coupling $J_3<0$ versus $J_1/|J_3|$.
 ``$120^\circ$" means the commensurate $120^\circ$ spin state with the three-sublattice periodicity, while ``incommensurate'' means the incommensurate spiral states.
\textcolor{black}{(b) The Fermi wavenumber $k_{\rm F}$ dependence of the ordering wavevector $|\bm{q}^*|$ around the incommensurate region of $2\pi/2.98 < k_{\rm F} \le 2\pi/2.58$, which is obtained from the Ewald RKKY potential for $L=32$.}
}
\label{inter}
\end{figure}

 Since the spin $\bm{S}_i$ is classical here, the ground state of the model in zero field can be obtained by the Fourier transform of $J_{ij}$, which yields the spiral order for certain ranges of $k_{\rm F}$ with the incommensurate ordering wavevectors. Fig. \ref{inter} (a) shows the ground-state spin structures of the model as compared with those of the short-range $J_1$-$J_3$ model of Ref. \cite{okubo2012multiple}. In addition to the ferromagnetic and the commensurate $120^\circ$-type orders, the incommensurate spiral order is realized for finite ranges of $k_{\rm F}$, similarly to the short-range $J_1$-$J_3$ case.  \textcolor{black}{We show in Fig.~\ref{inter}~(b) the Fermi wavenumber $k_{\rm F}$ dependence of the ordering wavenumber of the incommensurate spiral, $|\bm{q}^*|$, which is related to the sizes of skyrmions, i.e., the skyrmion diameter being given by $\frac{4\pi}{\sqrt{3}|\bm{q}^*|}$. As can be seen from Fig.~\ref{inter}~(b), the ordering wavenumbers cover a wide range of $|\bm{q}|$ spanning continuously from the short-wavelength $120^\circ$ structure to the long-wavelength uniform state.} In this study, we concentrate on the incommensurate case and mostly set $k_{\rm F} = 2\pi/2.71$. The ground-state wavenumbers are three-fold degenerate as $q_1^*, q_2^*$ and $q_3^*$, reflecting the three-fold rotational symmetry of the triangular lattice.

 Here, we comment on the relation of the present RKKY interaction (\ref{RKKY}) with the spin-quadratic interaction employed in the $q$-space Hamiltonian of Refs. \cite{hayami2017effective, hayami2019effect, hayami2020degeneracy, hayami2021noncoplanar, hayami2021square, yambe2021skyrmion}, i.e.,  $\sum_{\bm{q} = \bm{q}^*} J_{\bm{q}}|\bm{S}_{\bm{q}}|^2$, where the wavevectors $q_1^*\sim q_3^*$ are considered to be determined by the Fermi-surface effect (nesting). When such a quadratic interaction is rewritten in real space, one gets  $\sum_{i<j}\sum_{\bm{q} = \bm{q}^*} \cos(\bm{q}\cdot(\bm{r}_i - \bm{r}_j)) \bm{S}_i \cdot \bm{S}_j$. This quadratic interaction oscillates in sign depending on the distance, just as in the RKKY interaction (\ref{RKKY}), yielding {\it frustration\/} in the standard sense, like the frustration in metallic spin glasses \cite{mezard1987spin, mydosh1993spin, kawamura2015spin}. By contrast, it does not decay spatially even in the long-distance limit and corresponds to the mean-field approximation. Such an artificial feature arises due to the negligence of wavevector fluctuations around $q^*$. In the present study, we employ the ``textbook'' RKKY interaction, i.e., we take account of the spatial decay of the interaction caused by the wavevector fluctuations without assuming a specific Fermi-surface effect.

 The lattice sizes studied are $L=32,48,64$ with periodic boundary conditions. To fully take account of the long-range nature of the RKKY interaction beyond the finite system size $L$, we employ the Ewald sum method, which is a general method to treat the long-range interaction in numerical simulations \cite{ewald_p_p_1921_1424363, hansen1973statistical, fuchizaki1994towards, ikeda2008ordering, mitsumoto2021replica}. Details of the derivation of the Ewald potential for the RKKY interaction on the triangular lattice are given in Appendix~\ref{ewald_sum}.

 To equilibrate the system, we combine the Metropolis and over-relaxation methods. We employ the replica-exchange method at relatively high temperatures and the simulated annealing method at lower temperatures. We also employ the mixed-phase method \cite{creutz1979monte} to determine a part of the phase boundary. Our unit Monte Carlo step (MCS) consists of one Metropolis sweep and $L$ times over-relaxation sweeps. We take $1.0 \times 10^5$ MCS each for equilibration and computing thermal averages in the replica-exchange simulations. Errors of physical quantities are estimated over three independent runs.

\section{Monte Carlo results}\label{result}

\subsection{Phase diagram}\label{phase_diagram}

\begin{figure}[t]
\includegraphics[clip,width=85mm]{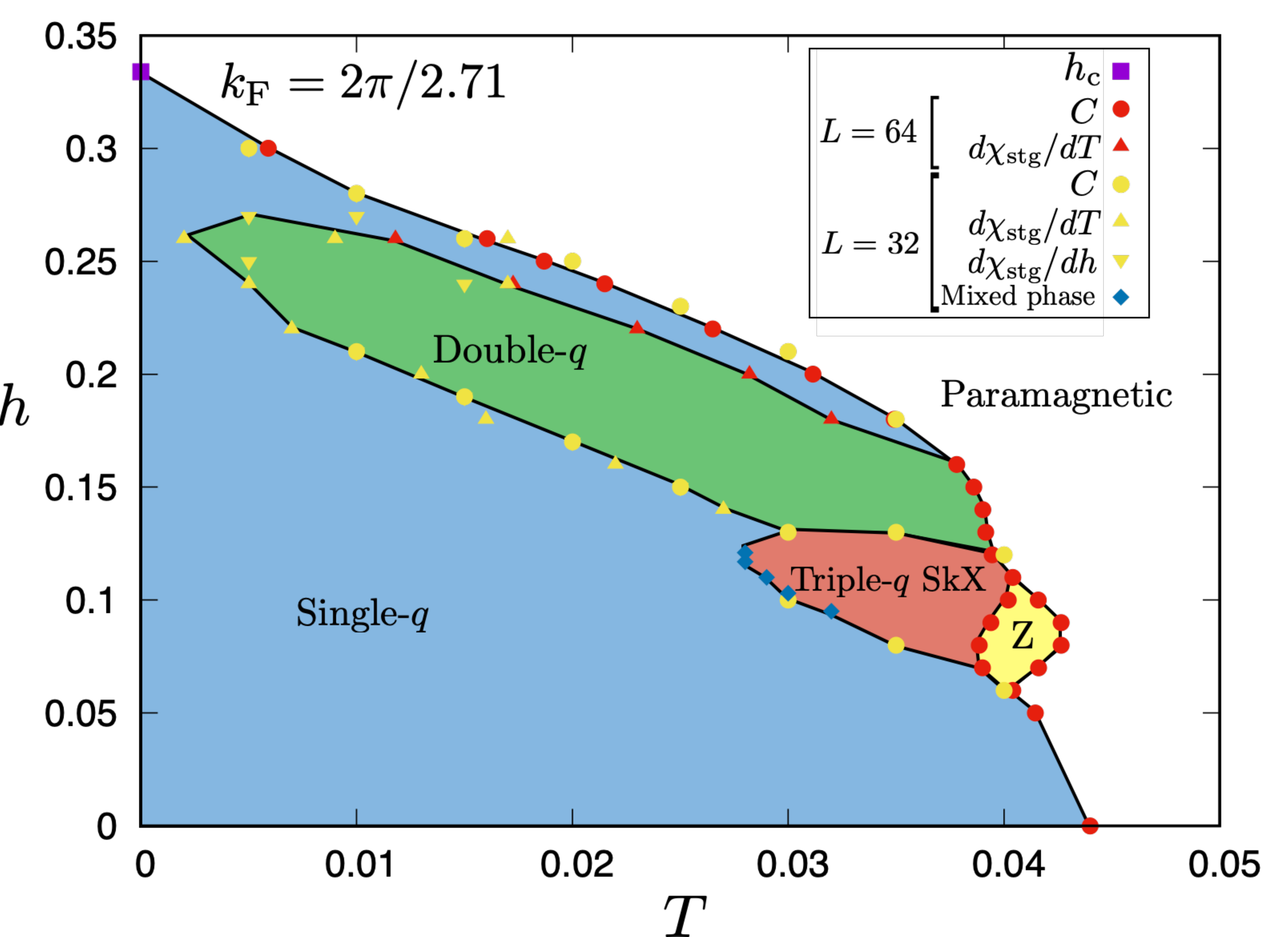}
\caption{Phase diagram of the 2D RKKY Heisenberg model on the triangular lattice with $k_{\rm F} = 2\pi/2.71$ in the temperature versus magnetic-field plane. Phase boundaries are determined from the peak positions of the specific heat and the $T$-derivative of the staggered scalar chirality, whereas the $T=0$ transition point $h_{\rm c}$ is determined from the analytic result of the ground-state energy. A part of the phase boundary is determined by the mixed-phase method.
}
\label{phase}
\end{figure}

In this section, we present our simulation results.
We first show in Fig. \ref{phase} the temperature $T$ versus the magnetic-field $h$ phase diagram of the model with $k_{\rm F} = 2\pi/2.71$. The triple-$q$ SkX phase is stabilized at finite temperatures and intermediate fields, in addition to the single-$q$, double-$q$, and $Z$ phases. The obtained phase diagram is similar to the one of the short-range $J_1$-$J_3$ model \cite{okubo2012multiple}, including the properties of each phase as will be shown in the following subsections and sections.

\subsection{Spin configurations in the wavevector- and real-spaces}\label{structure_factor}
\begin{figure}[t]
\includegraphics[clip,width=85mm]{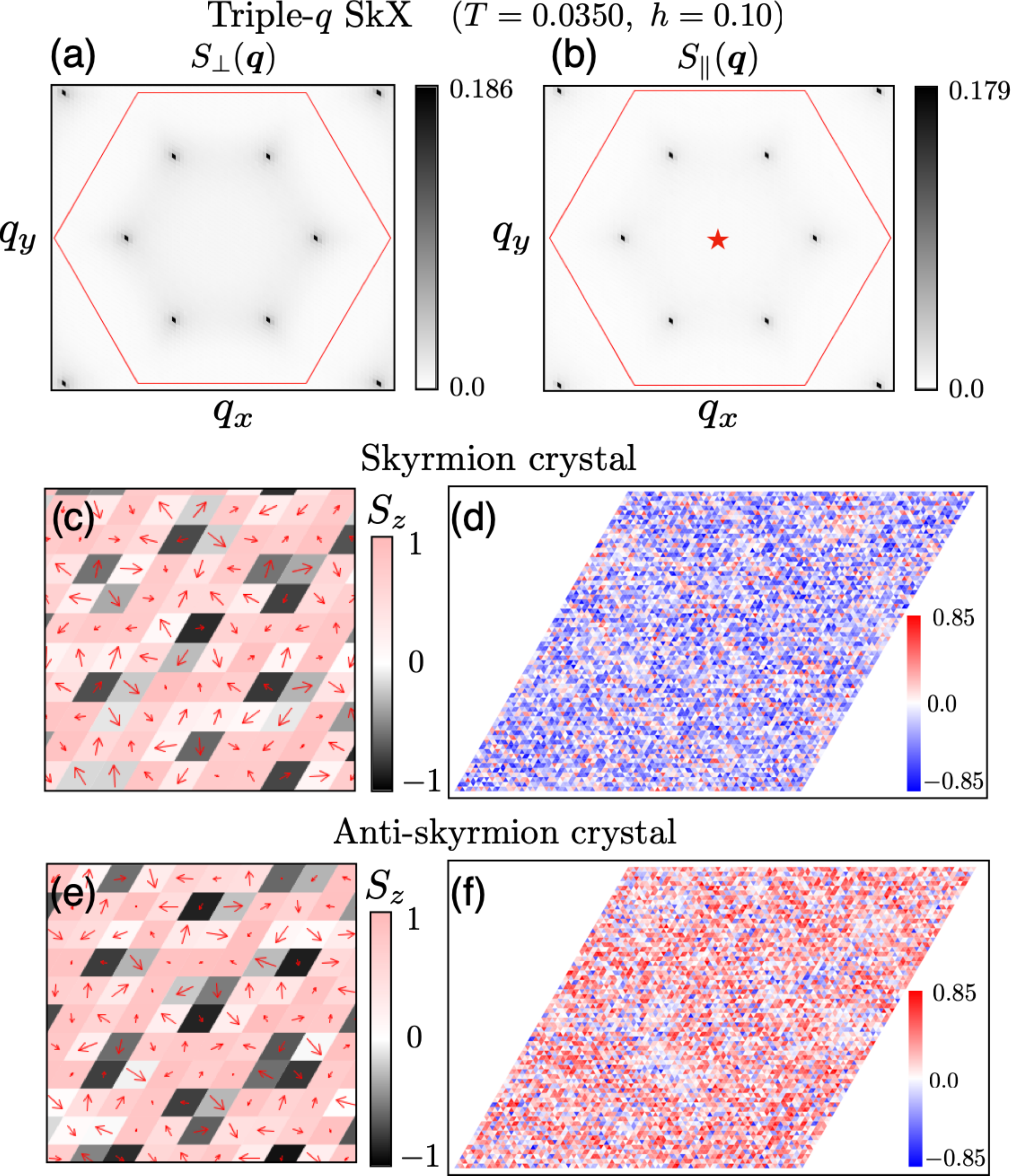}
\caption{
The spin configurations in the wavevector- and real-space spaces in the single-$q$ phase at $T=0.0350$ and $h=0.10$ for $k_{\rm F} = 2\pi/2.71$ and $L=64$. \textcolor{black}{(a) The spin structure factor for the spin $xy$-components $S_{\perp}(\bm{q})$, (b) the one for the spin $z$-component $S_{\parallel}(\bm{q})$,} (c, e) the real-space spin configurations corresponding to (c) the SkX and (d) the anti-SkX, and (d, f) the real-space chirality configurations corresponding to (e) the SkX and (f) the anti-SkX. In (a) and (b), the red hexagon represents the first Brillouin zone and the red star at the origin represents the intensive uniform $\bm{q} = \bm{0}$ component induced by applied magnetic fields. The structure factors are averaged over $10^4$ Monte Carlo steps with only the Metropolis updates. In (c-f), the short-time averaging over 100 MCS with only the Metropolis updating is made to reduce the thermal noise. The spin $xy$ components are represented by the arrow and the $z$ component by the colors of the rhombuses. The sign of the scalar chirality is represented by the colors of the triangles.
}
\label{triple_q}
\end{figure}
\begin{figure}[t]
\includegraphics[clip,width=85mm]{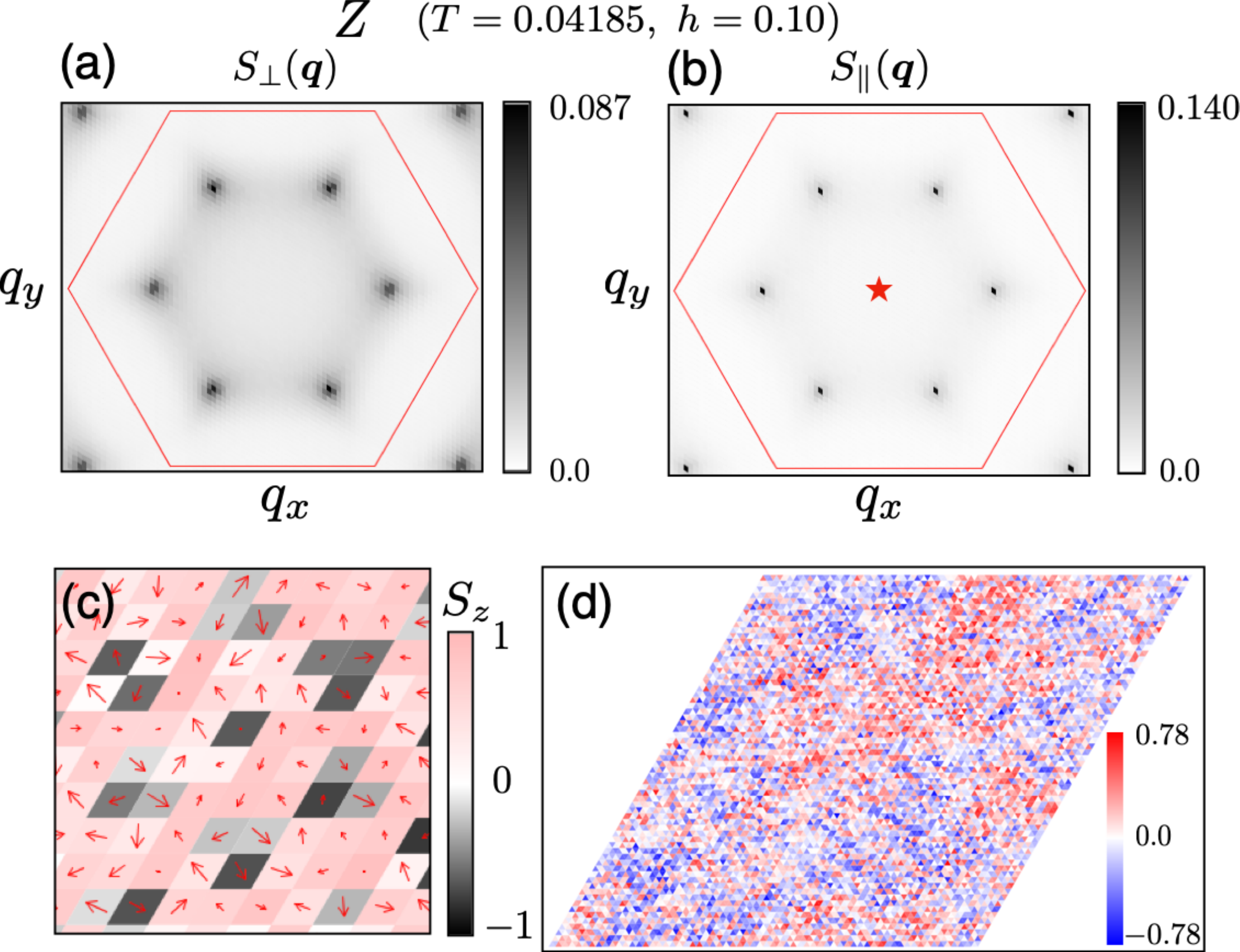}
\caption{
The spin configurations in the wavevector- and real-space spaces in the $Z$ phase at $T=0.04185$ and $h=0.10$ for  $k_{\rm F} = 2\pi/2.71$ and $L=64$. \textcolor{black}{(a) The spin structure factor for the spin $xy$-components $S_{\perp}(\bm{q})$, (b) the one for the spin $z$-component $S_{\parallel}(\bm{q})$,} (c) the real-space spin configuration, and (d) the real-space chirality configuration. The condition of the calculation and the meaning of the colors and symbols are the same as those of Fig. \ref{triple_q}.
}
\label{Z_phase}
\end{figure}
\begin{figure}[t]
\includegraphics[clip,width=85mm]{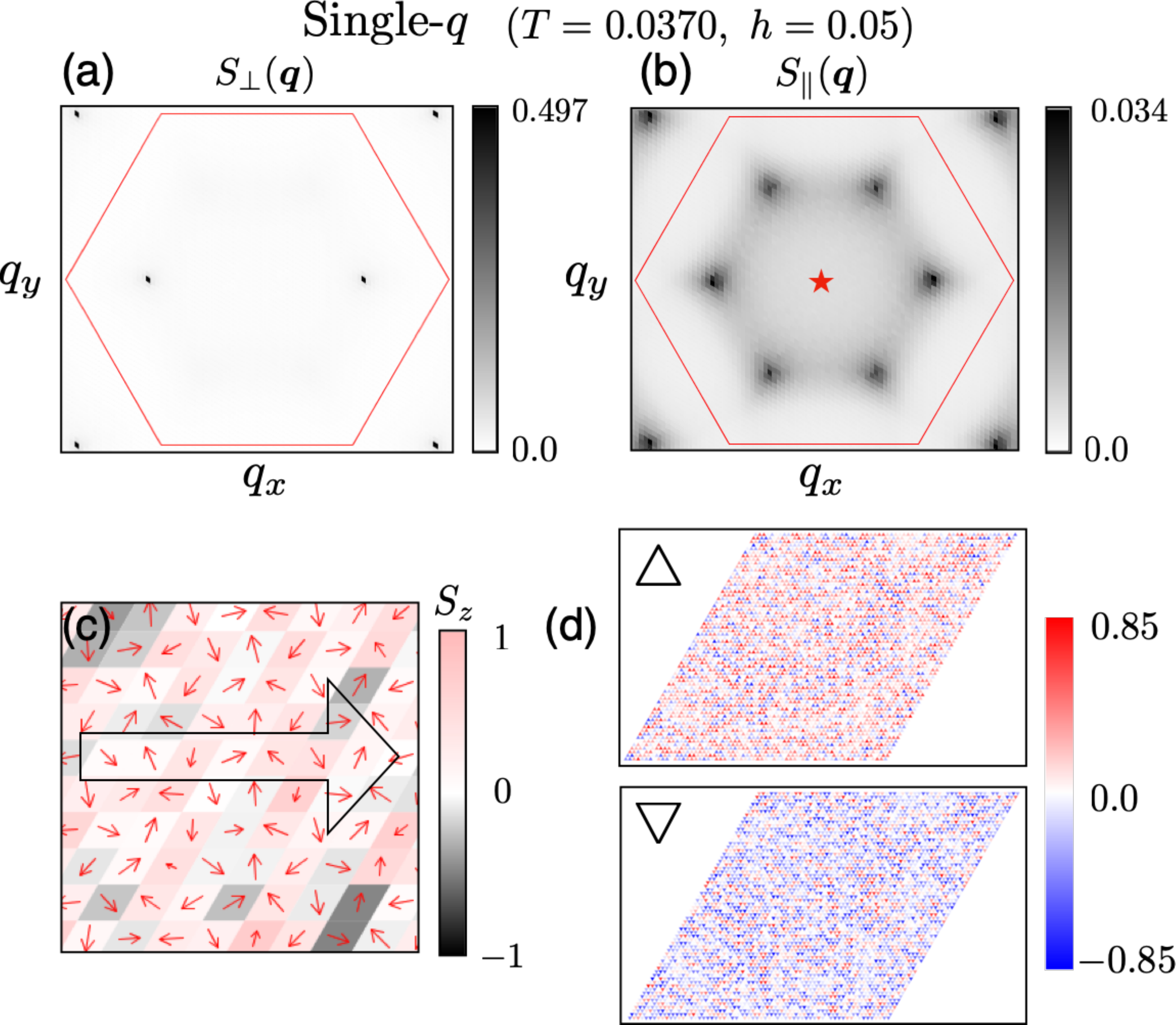}
\caption{
\textcolor{black}{The spin configurations in the wavevector- and real-space spaces in the single-$q$ phase at $T=0.0370$ and $h=0.05$ for  $k_{\rm F} = 2\pi/2.71$ and $L=64$. (a) The spin structure factor for the spin $xy$-components $S_{\perp}(\bm{q})$, (b) the one for the spin $z$-component $S_{\parallel}(\bm{q})$, (c) the real-space spin configuration, and (d) the real-space chirality configuration. The condition of the calculation and the meaning of the colors and symbols are the same as those of Fig. \ref{triple_q}.}
}
\label{single_q}
\end{figure}
\begin{figure}[t]
\includegraphics[clip,width=85mm]{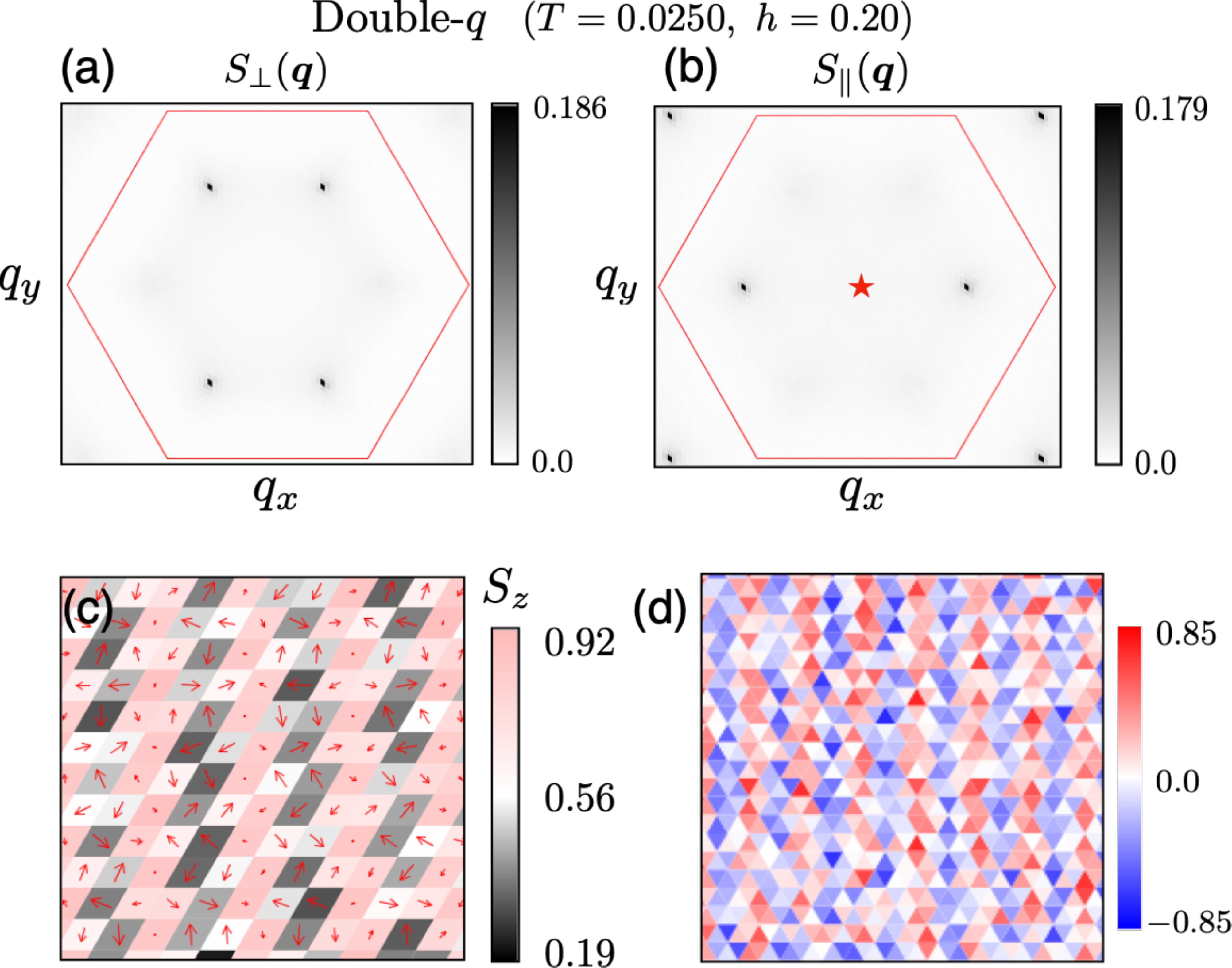}
\caption{
\textcolor{black}{The spin configurations in the wavevector- and real-space spaces in the double-$q$ phase at $T=0.0250$ and $h=0.20$ for  $k_{\rm F} = 2\pi/2.71$ and $L=64$. (a) The spin structure factor for the spin $xy$-components $S_{\perp}(\bm{q})$, (b) the one for the spin $z$-component $S_{\parallel}(\bm{q})$, (c) the real-space spin configuration, and (d) the real-space chirality configuration. The condition of the calculation and the meaning of the colors and symbols are the same as those of Fig. \ref{triple_q}.}
}
\label{double_q}
\end{figure}
%
%

%
%

\textcolor{black}{Next, we wish to characterize each phase appearing in the phase diagram of Fig.\ref{phase} by presenting our MC results of the spin structure factors, and the spin and chirality configurations in real space. The spin structure factor for the spin $xy$-components $S_{\perp}(\bm{q})$ and the one for the spin $z$-component $S_{\parallel}(\bm{q})$ are defined by
\begin{eqnarray}
S_\perp (\bm{q}) &=& \frac{1}{N} \left\langle \left( \sum_{\mu = x,y} \left| \sum_{i=1}^N S_i^\mu e^{-i \bm{q}\cdot \bm{r}_i}\right|^2 \right)^{1/2} \right\rangle, \label{insta_perp} \\
S_\parallel (\bm{q}) &=& \frac{1}{N} \left\langle \left| \sum_{i=1}^N S_i^z e^{-i \bm{q}\cdot \bm{r}_i} \right| \right\rangle, \label{insta_para}
\end{eqnarray}
where $\langle \cdots \rangle$ represents the thermal average.}

 \textcolor{black}{The spin structure factors $S_\perp (\bm{q})$ and $S_\parallel (\bm{q})$ computed in the triple-$q$ SkX phase are shown in  Figs. \ref{triple_q} (a) and (b), respectively. In the triple-$q$ SkX phase, both the spin $xy$-components and $z$-component are characterized by all three equivalent pairs of ordering wavevectors, $\pm \bm{q}_1^*,~\pm \bm{q}_2^*$ and $\pm \bm{q}_3^*$, keeping the $Z_3$ lattice-rotation symemtry. Note that, due to the $U(1)$ spin-rotation symmetry associated with the spin $xy$-components, sharp-looking peaks in $S_\perp (\bm{q})$ might not be true Bragg peaks but rather quasi-Bragg peaks associated with algebraic spin correlations \cite{mermin1966absence, bruno2001absence}. }

 Spin and chirality configurations in real space in the triple-$q$ SkX phase are shown in Figs.\ref{triple_q} (c)-(f). The local scalar chirality is defined for the upward (downward) triangles by $\chi_{\bigtriangleup(\bigtriangledown)} = \bm{S}_i \cdot \bm{S}_j \times \bm{S}_k~(i,j,k \in \bigtriangleup(\bigtriangledown))$. Due to the underlying $Z_2$ spin-reflection symmetry, the triple-$q$ SkX and anti-SkX states are energetically degenerate, either of which is realized as a result of the spontaneous $Z_2$-symemtry breaking, as shown in the spin configurations of Figs.~\ref{triple_q}~(c,e) and the chirality configurations of Figs.~\ref{triple_q}~(d,f). The SkX and anti-SkX states have net total chiralities of mutually opposite signs, giving rise to the topological Hall effect of mutually opposite signs. 

 \textcolor{black}{The spin structure factors of the $Z$ phase are shown in Figs. \ref{Z_phase} (a) and (b). While they look more or less similar to those of the triple-$q$ SkX phase, the intensities in $S_\perp (\bm{q})$ are rather broad with finite widths, not being even the quasi-Bragg peaks, in sharp contrast to the intensities in $S_\parallel (\bm{q})$.} \textcolor{black}{Such sharp versus broad features in $S_\parallel({\bm q})$ and  $S_\perp({\bm q})$ intensities might serve as an experimental indicator of the $Z$ phase in distinction with the triple-$q$ SkX phase where both $S_\parallel({\bm q})$ and  $S_\perp({\bm q})$ exhibit sharp intensities.}

 As shown in the real-space spin and chirality configurations shown in Figs. \ref{Z_phase} (c) and (d), \textcolor{black}{the $Z$ phase is the random-domain state consisting of the SkX and anti-SkX domains. On lowering $T$ toward the transition point to the triple-$q$ SkX phase, the sizes of these random domains grow, eventually leading to the triple-$q$ SkX state characterized by the net total chirality. Such behaviors of the spin structure factors and the real-space spin and chirality configurations are quite similar to those found in the short-range $J_1$-$J_3$ model \cite{okubo2012multiple}. Note that in both cases of the short-range and long-range models the finite-size SkX and anti-SkX domains in the $Z$ phase would be subject to thermal fluctuations, while more macroscopic SkX and anti-SkX domains in the triple-$q$ phase would be more or less frozen. Such a difference might be distinguishable by appropriate experimental means, e.g., by Lorentz Transmission Electron Microscopy.}

 \textcolor{black}{The spin structure factors and the real-space spin and chirality configurations in the single-$q$ phase are shown in Figs. \ref{single_q}~(a,b) and \ref{single_q}~(c,d), respectively. $S_{\perp}(\bm{q})$ is characterized by one pair of ordering wavevectors selected from three equivalent wavevector pairs as shown in Fig. \ref{single_q} (a), spontaneously breaking the lattice-rotation symmetry, which corresponds to the spiral direction in real space as shown in Fig \ref{single_q} (c). By contrast, $S_{\parallel}(\bm{q})$ exhibits only weak broad intensities at $\pm \bm{q}_1$, $\pm \bm{q}_2$ and $\pm \bm{q}_3$. As shown in Fig.\ref{single_q} (d), the scalar chirality in the single-$q$ phase exhibits a staggered order, i.e., the upward and downward triangles have mutually opposite signs of the local scalar chirality.}

 \textcolor{black}{The spin structure factors and the real-space spin and chirality configurations in the double-$q$ phase are shown in Figs.~\ref{double_q}~(a,b) and \ref{double_q}~(c,d), respectively. $S_{\perp}(\bm{q})$ in the double-$q$ phase is characterized by two pairs of ordering wavevectors. The state consists of the superposition of the two-spirals formed by the spin $xy$-components running along different directions as shown in the real-space spin configuration of Fig. \ref{double_q} (c). By contrast, the spin $z$-component forms the linear spin-density wave running along the direction complementary to those of the $xy$-components, and the corresponding one-pair peak is observed in $S_\parallel (\bm{q})$ as shown in Fig. \ref{double_q} (b). The scalar chirality shown in Fig. \ref{double_q} (d) also exhibits a linear density wave in the same direction as that of the spin $z$-component.}

\subsection{Temperature dependences of several physical quantities}\label{phys_quantities}

\begin{figure}[t]
\includegraphics[clip,width=85mm]{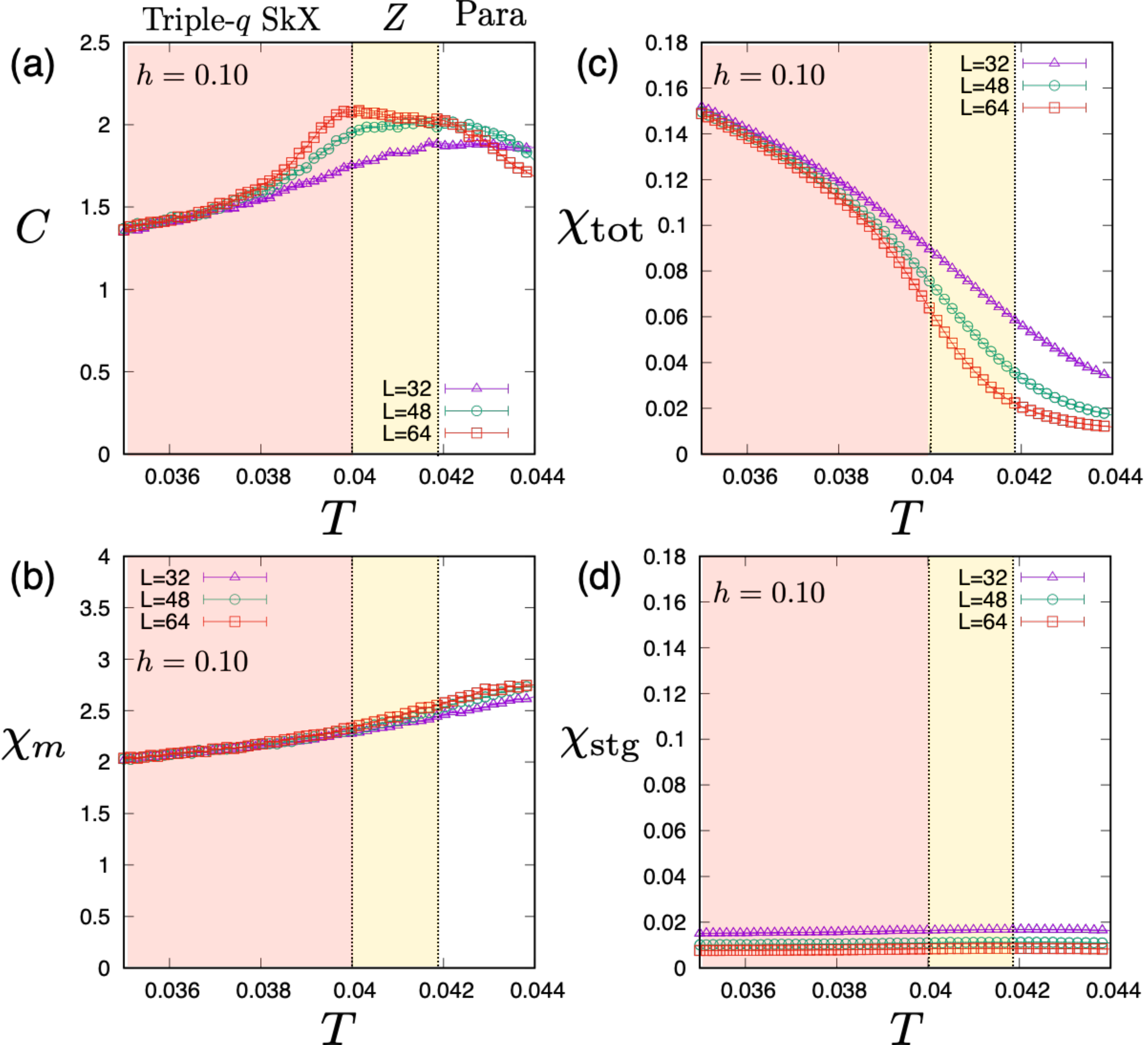}
\caption{
\textcolor{black}{The temperature dependences of (a) the specific heat, (b) the uniform susceptibility, (c) the total scalar chirality, (d) the staggered scalar chirality at the intermediate magnetic field of $h=0.10$, crossing in the phase diagram the paramagnetic, $Z$, and triple-$q$ SkX phases.}
}
\label{tdep_mid}
\end{figure}
\begin{figure}[t]
\includegraphics[clip,width=85mm]{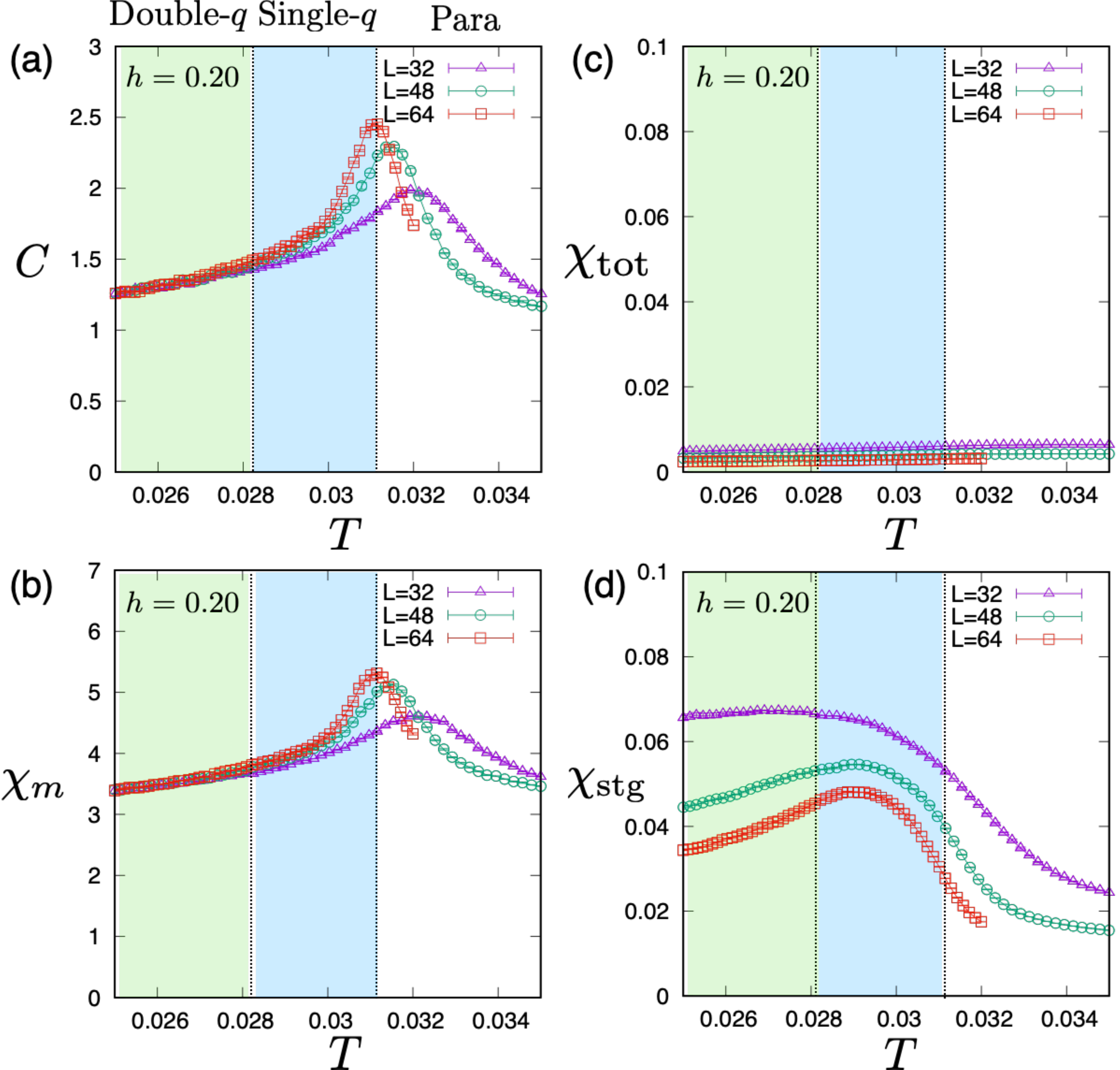}
\caption{
\textcolor{black}{The temperature dependences of (a) the specific heat, (b) the uniform susceptibility, (c) the total scalar chirality, (d) the staggered scalar chirality at the high magnetic field of $h=0.20$, crossing in the phase diagram the paramagnetic, single-$q$, and double-$q$ phases.}
}
\label{tdep_high}
\end{figure}

\textcolor{black}{To further clarify the properties of each phase, we next investigate the temperature dependences of several physical quantities, including the specific heat $C$, the uniform magnetic susceptibility $\chi_m$, the total scalar chirality $\chi_{\rm tot}$ and the staggered scalar chirality $\chi_{\rm stg}$. The specific heat and the susceptibility are defined by the energy and the magnetization fluctuations, respectively. 
The total and staggerd scalar chiralities, which can be regarded as the order parameters of the triple-$q$ SkX and single-$q$ phases, respectively, are defined by
\begin{align}
\chi_{\rm tot} &= \frac{1}{2N}\left(\sum_{\bigtriangleup} \chi_{\bigtriangleup} + \sum_{\bigtriangledown} \chi_{\bigtriangledown}\right), \\
\chi_{\rm stg} &= \frac{1}{2N}\left(\sum_{\bigtriangleup} \chi_{\bigtriangleup} - \sum_{\bigtriangledown} \chi_{\bigtriangledown}\right), 
\end{align}
where the summation $\sum_\bigtriangleup$ ($\sum_\bigtriangledown$) runs over all upward (downward) triangles on the triangular lattice. 
In the following, we show our MC data obtained by fully equilibrated replica-exchange simulations at two typical magnetic fields, i.e., the intermediate field of $h=0.10$ crossing in the phase diagram the paramagnetic, $Z$, and triple-$q$ SkX phases, and the high field of $h=0.20$ crossing in the phase diagram the paramagnetic, single-$q$, and double-$q$ phases. }

 \textcolor{black}{Figs.~\ref{tdep_mid}~(a)-(d) shows the data at the intermediate magnetic field of $h=0.10$ in the temperature range of $T\geq 0.035$. As can be seen from Fig. \ref{tdep_mid} (a), the specific heat exhibits two peaks at $T_{\rm c1}^{\rm (mid)}=0.0419$ and at $T_{\rm c2}^{\rm (mid)}=0.0400$. As shown in Fig.~\ref{tdep_mid}~(c), the total scalar chirality $\chi_{\rm tot}$ is suppressed at $T_{\rm c2}^{\rm (mid)} < T <T_{\rm c1}^{\rm (mid)}$ but is enhanced at $T<T_{c2}^{\rm(mid)}$. By contrast, the staggered scalar chirality $\chi_{\rm stg}$ shown in Fig.~\ref{tdep_mid}~(d) is vanishing in the investigated temperature range. Such behaviors are consistent with the identification of $T_{\rm c1}^{\rm (mid)}$ and $T_{\rm c2}^{\rm (mid)}$ as the paramagnetic/$Z$ and the $Z$/triple-$q$ transition temperatures, respectively. Note that, on further lowering $T$ to $T\leq 0.035$, the system exhibits another phase transition into the single-$q$ phase, as shown in the phase diagram of Fig.~\ref{phase}.}

 \textcolor{black}{Figs.~\ref{tdep_high}~(a)-(d) shows the data at the high magnetic field of $h=0.20$ in the temperature range of $T\geq 0.025$. Although the specific field shown in Fig. \ref{tdep_high} (a) exhibits only a single peak at $T_{\rm c1}^{\rm (high)}=0.0312$, the staggered scalar chirality $\chi_{\rm stg}$ shown in Fig. \ref{tdep_high} (d) is enhanced in the intermediate temperature range $T_{\rm c2}^{\rm (high)} = 0.0282<T<T_{\rm c1}^{\rm (high)}$, but is suppressed below $T_{\rm c2}^{\rm (high)}$. In addition, the behaviors of the $Z_3$-symmetry-breaking parameters both for the spin $xy$ and $z$ components, shown in Appendix \ref{rho3} for each typical magnetic field, suggest that the $Z_3$ lattice-rotation symmetry is broken only for the spin $xy$ components but not for the $z$ component at $T_{\rm c2}^{\rm (high)}<T<T_{\rm c1}^{\rm (high)}$, while it is broken both for the spin $xy$ and $z$ components at $T<T_{\rm c2}^{\rm (high)}$. The total scalar chirality $\chi_{\rm tot}$ shown in Fig. \ref{tdep_high} (c) tends to vanish at any temperature. These observations are consistent with the identification of $T_{\rm c1}^{\rm (high)}$ and $T_{\rm c2}^{\rm (high)}$ as the paramagnetic/single-$q$ and the single-$q$/double-$q$ transition temperatures, respectively. The magnetic susceptibility $\chi_{m}$ shown in Fig.~\ref{tdep_high}~(b) exhibits a behavior more or less similar to the specific heat. Note that, on further lowering the temperature to $T\leq 0.025$, the system exhibits another phase transition into the single-$q$ phase, as shown in the phase diagram of Fig.~\ref{phase}.}

 \textcolor{black}{The corresponding data at the low field of $h=0.05$, crossing in the phase diagram the paramagnetic and single-$q$ phases, are shown in Appendix~\ref{low_field}. }

\subsection{Distribution functions of the chirality}\label{absence}

\begin{figure}[t]
\includegraphics[clip,width=85mm]{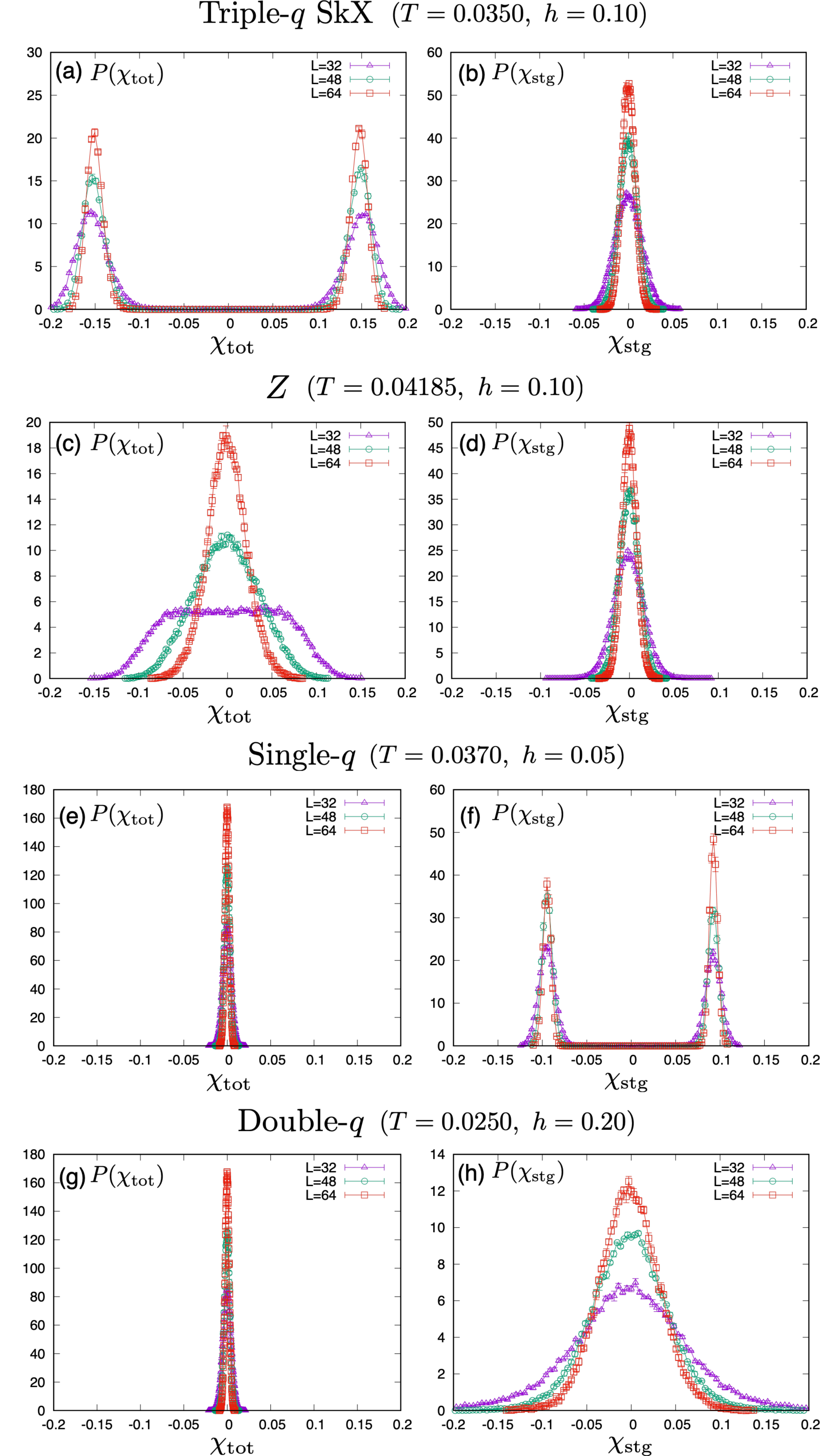}
\caption{Distribution functions of the total scalar chirality $\chi_{\rm tot}$ [left column] and the staggered scalar chirality $\chi_{\rm stg}$ [right column] for various sizes $L$.
(a) and (b) are obtained in the triple-$q$ SkX phase, \textcolor{black}{(c) and (d) in the $Z$ phase, (e) and (f) in the single-$q$ phase, (g) and (h) in the double-$q$ phase. }
 The distribution function is normalized as $\int_{-\infty}^{\infty} P(\chi)d\chi =1$.
}
\label{hist}
\end{figure}

 Very recently, an intriguing RSB phenomenon was observed in the triple-$q$ SkX phase and the double-$q$ phase in the 3D RKKY Heisenberg model on a stacked-triangular lattice where the multiple-$q$ states macroscopically coexist with the single-$q$ state \cite{mitsumoto2021replica}. Under such circumstances, we wish to examine in this subsection whether a similar RSB occurs or not in the present 2D RKKY model. For this purpose, we study the distribution functions of the total and scalar chiralities in each phase.

 In Fig. \ref{hist} (a), we show the distribution function of the total scalar chirality $P(\chi_{\rm tot})$ at $T=0.035$ and $h=0.1$ lying in the triple-$q$ SkX phase. Reflecting the $Z_2$ chiral degeneracy, there are positive and negative peaks of equal heights, each corresponding to the anti-SkX and SkX states, respectively, which grow and sharpen with $L$. By contrast, no appreciable weight is found around $\chi_{\rm tot}=0$ corresponding to the single-$q$ state, in sharp contrast to the 3D case. Hence, no signature of the single-$q$ state coexisting with the triple-$q$ SkX (anti-SkX) state is observed, suggesting the absence of the RSB. Noting that the order parameter of the single-$q$ state is the staggered scalar chirality $\chi_{\rm stag}$, we show in Fig. \ref{hist} (b) its distribution function $P(\chi_{\rm stag})$. Only a sharp peak around zero corresponding to the triple-$q$ state is observed without any sign of the single-$q$ state, again indicating the absence of the RSB. 

 \textcolor{black}{Likewise, there is no sign of the RSB in other phases, whose distribution functions of the total and staggered chiralities are shown in Figs. \ref{hist} (c)-(h). The staggered scalar chirality becomes finite only in the single-$q$ phase, and except in the single-$q$ phase itself, $P(\chi_{\rm stg})$ exhibits only a single peak around $\chi_{\rm stg}=0$, indicating that the single-$q$ state coexists with neither the double-$q$ state nor the $Z$ state.}

 \textcolor{black}{We also present the distribution functions of the $Z_3$ symmetry breaking order parameter in Appendix \ref{histogram}. The distributions of the $Z_3$ symmetry breaking order parameter also indicate that the RSB does not occur in the 2D RKKY model. }

\section{Summary and Discussion}\label{summary}

 In summary, we have shown by extensive MC simulations on the 2D RKKY model without nesting, four-body interaction or magnetic anisotropy that the SkX (anti-SkX) phase is stabilized under magnetic fields only by the isotropic RKKY interaction. The SkX is stabilized by thermal fluctuations and frustration inherent to the RKKY interactions decaying as $1/r^3$ with the alternating signs. We have also shown the RSB does not occur in the 2D RKKY model, in sharp contrast to the 3D RKKY model.

 \textcolor{black}{We wish to discuss first why the RSB does not occur in the present 2D RKKY model, unlike the corresponding 3D RKKY model. In the fields of spin glasses and structural glasses where the RSB was first introduced and discussed, understanding of the RSB in finite dimensions has been regarded as a very hard and challenging problem and is still debated \cite{binder1986spin, mezard1987spin, kawamura2015spin, parisi_urbani_zamponi_2020}. In the case of glassy systems, while the occurrence of the RSB has been established in infinite dimensions, the situation is not necessarily clear in finite dimensions, though the general expectation might be that the RSB is more likely to arise in higher dimensions. Such a general tendency seems consistent with our present observation that the RSB is realized in the 3D RKKY system but not in the 2D RKKY system.}

 \textcolor{black}{Here, we wish to give a further discussion on the issue of the existence/nonexistence of the RSB in the present RKKY model from the viewpoint of the underlying {\it frustration\/} effect, a crucial ingredient to realize the SkX state. In the 3D RKKY model studied in Ref.\cite{mitsumoto2021replica}, not only the in-plane interaction but also the inter-plane interaction is frustrated due to the oscillating nature of the RKKY interaction. Although the ground state of the 3D RKKY model studied in Ref.\cite{mitsumoto2021replica} exhibits the ferromagnetic alignment along the inter-plane direction, apparently disguising the absence of inter-plane frustration, the ferromagnetic inter-plane spin alignment actually frustrates with the RKKY interaction which oscillates in sign, so that the 3D RKKY model is frustrated not only in the in-plane directions but also in the inter-plane direction. In that sense, the 3D RKKY model is more heavily frustrated than the 2D RKKY model, the latter possessing only the in-plane frustration.}

 \textcolor{black}{The condition of the RSB is that the free energy difference between the macroscopically coexisting phases, which are the triple-$q$ SkX state and the single-$q$ state in the 3D RKKY model, stays only of $O(1)$ even in the thermodynamic limit. Such an unusual situation is expected to arise only in the presence of quite heavy frustration. Thus, the absence/presence of the RSB in the 2D/3D RKKY model might intuitively be understandable from the difference in the severity of the underlying frustration effect between the 2D/3D RKKY models. }

 \textcolor{black}{In this connection, if one considers the 3D version of the short-range $J_1$-$J_3$ model with the ferromagnetic inter-plane coupling, the SkX phase there is expected not to exhibit the RSB since the ferromagnetic inter-plane interaction does not bring about any additional frustration along the inter-plane direction. Indeed, the ordering properties of such a 3D model with the short-range interactions have been studied quite recently, indicating that the triple-$q$ SkX phase is realized, but not accompanying the RSB even in 3D \cite{osamura}. This observation highlights and confirms the importance of the severe frustration effect caused by the inter-plane RKKY interaction in realizing the RSB phenomenon.}

 
Next, we discuss the possible implications of our results to the SkX formation mechanism. The nesting property of the Fermi surface, which has been supposed to be important in the SkX formation in previous studies \cite{ozawa2017zero, hayami2017effective, hayami2020degeneracy}, might play a role in picking up particular wavevectors, but frustration alone can pick up appropriate ordering wavevectors even without nesting as our present model calculation has shown. In that sense, nesting plays only a secondary role in the SkX formation itself. As the system without nesting can exhibit a stable SkX phase, and the tuning of the interaction is usually necessary to realize the strong nesting, it would be important to check experimentally whether the candidate material really possesses strong nesting or not by independent measurements, e.g., the angle-resolved photoemission spectroscopy \cite{inosov2009electronic, takahashi2020competing}. 

 As discussed in Ref. \cite{okubo2012multiple} at the Ginzburg-Landau level, the effect of emergent many-body interactions generated by thermal effects is important in stabilizing the SkX state. In fact, as shown in Appendix \ref{evaluation}, such emergent many-body interactions can rigorously be derived in the form of the expansion in terms of the coarse-grained effective fields so long as $T$ is not too low compared with $T_c$ \cite{fisher1983scaling, kawamura1988renormalization}. We note that these many-body terms, including the biquadratic interaction with the positive coefficient, have the same forms as those discussed in Ref.\cite{hayami2017effective} in the context of the electronic perturbation with respect to $J_{sd}/\epsilon_{\rm F}$. Note that while the energy scale of the four-body (biquadratic) interaction of Ref.\cite{hayami2017effective} is smaller than that of the quadratic RKKY interaction $J_0$ by the factor of $(J_{sd}/\epsilon_{\rm F})^2$, the energy scale of the emergent four-body interaction arising from thermal fluctuations is of the same order as $J_0$, not containing the small parameter $(J_{sd}/\epsilon_{\rm F})^2$. With Gd$_2$PdSi$_3$ in mind, we estimate from the available experimental data the typical $(J_{sd}/\epsilon_{\rm F})^2$ value to be as small as $\sim 10^{-3}-10^{-4}$: Details are given in Appendix F. Hence, at least at finite $T$, the emergent many-body interaction originating from thermal fluctuations would play a dominant role in stabilizing the SkX.
 
Interesting questions are whether the SkX is stabilized at $T\rightarrow 0$ and its stabilization mechanism. \textcolor{black}{It has been reported experimentally that the SkX state, once stabilized at finite temperatures, can be continued down to lower temperatures as a metastable state \cite{kagawa2015, kagawa2016}}. One possible mechanism to realize the stable $T=0$ SkX state might be the biquadratic interaction of electronic origin as discussed in Refs.\cite{hayami2017effective}. We note that {\it quantum fluctuations\/} of localized spins often play similar roles as thermal fluctuations in frustrated magnets, stabilizing the nontrivial multiple-$q$ state even at $T=0$ even when the classical ground state is of the single-$q$ type \cite{marmorini2014magnon}. Again, such a term originating from quantum spin fluctuations does not contain the small parameter $(J_{sd}/\epsilon_{\rm F})^2$ and might play a role in the SkX stabilization at lower $T$, especially when the spin quantum number $S$ is small. Although we limit our discussion so far to isotropic Heisenberg spins, real magnets possess some amount of magnetic anisotropy. Indeed, previous theoretical studies have indicated that the easy-axis-type anisotropy tends to stabilize the SkX state even at $T=0$ both for the classical \cite{leonov2015multiply, hayami2016bubble, wang2020skyrmion, wang2021meron} and \textcolor{black}{quantum \cite{lohani2019quantum} systems.} Thus, the magnetic anisotropy might actually play a major role in the SkX stabilization at $T=0$.

\begin{acknowledgments}
The authors would like to thank K. Aoyama for useful discussion. We are thankful to ISSP, the University of Tokyo, and YITP, Kyoto University, for providing us with CPU time. This work is supported by JSPS KAKENHI Grant No. JP17H06137.
\end{acknowledgments}

\appendix

\section{Ewald-sum method of the RKKY interaction on the two-dimensional triangular lattice}\label{ewald_sum}

In this section,  we explain some of the details of the application of the Ewald-sum method, widely used in computing the long-range interactions in periodic systems of finite sizes \cite{ewald_p_p_1921_1424363, hansen1973statistical, fuchizaki1994towards, ikeda2008ordering, mitsumoto2021replica}, to the RKKY Heisenberg model on the 2D triangular lattice. In fact, in the straightforward application of the Ewald method to the present 2D problem, we meet some computational difficulty due to poor numerical convergence. In order to circumvent this difficulty, we first consider the 3D system and then take the 2D limit, i.e., the limit of the ratio $c$ between the interlayer and intralayer distances tending to infinity at the end. 

 We begin with the Ewald potential for the RKKY interaction on the 3D stacked-triangular lattice employed in Ref.\cite{mitsumoto2021replica},
\begin{eqnarray}
J_{ij}^{\rm Ewald(3D)} &=& -J_0\sum_{\bm{\lambda}} J_{ij}(\bm{ \lambda}), \label{ewald} \\
J_{ij}(\bm{\lambda}) &=& \frac{\cos (2k_{\rm F} |\bm{r}_{ij} + L \bm{\lambda}|)}{|\bm{r}_{ij} + L \bm{\lambda}|^3} \nonumber \\
&-& \frac{\sin (2k_{\rm F} |\bm{r}_{ij} + L \bm{\lambda}|)}{2k_{\rm F}|\bm{r}_{ij} + L \bm{\lambda}|^4},
\end{eqnarray}
where $L$ is linear size of the system, and $\bm{\lambda} = n_a \bm{a} + n_b \bm{b} + n_c \bm{c}$ with $\bm{a} = (1,0,0),~\bm{b} = (1/2,\sqrt{3}/2,0),~\bm{c} = (0,0,c)$. The sum over $\bm{\lambda}$ in Eq. (\ref{ewald}) runs over integers $n_\mu = -\infty,...,0,...\infty$ ($\mu = a,b,c)$, $L \bm{\lambda}$ mapping the original cell of the size $L^3$ to the image cell with exactly the same spin configuration as that in the original cell.

We divide the potential into two parts, a short-range contribution and a long-range contribution, by inserting the identity,
\begin{equation}
1 = \frac{1}{\Gamma (\frac{3}{2})} \left[ \Gamma \left( \frac{3}{2},\pi\frac{|\bm{r}_{ij} + L \bm{\lambda}|^2}{L^2} \right) + \gamma \left( \frac{3}{2},\pi\frac{|\bm{r}_{ij} + L \bm{\lambda}|^2}{L^2}\right) \right],
\end{equation}
where $\Gamma(\alpha) = \int_0^\infty dt e^{-t} t^{\alpha -1}$ is the gamma function, $\Gamma(\alpha, x) = \int_x^\infty dt e^{-t} t^{\alpha -1}$ and $\gamma(\alpha, x) = \int_0^x dt e^{-t} t^{\alpha -1}$ are the incomplete gamma functions.
The short-range contribution is well-converged in the real space. To also make the long-range contribution well-converged, we transform it into the Fourier space. Then, Eq. (\ref{ewald}) can be rewritten as \cite{mitsumoto2021replica}, 
\begin{widetext}
\begin{eqnarray}
J_{ij}^{\rm Ewald(3D)} &=& -\frac{J_0}{\Gamma(\frac{3}{2})} \left(J_{ij}^{\rm short(3D)} + J_{ij}^{\rm long(3D)} \right) \\
J_{ij}^{\rm short(3D)} &=& \sum_{\bm{\lambda}} \Gamma \left(\frac{3}{2},\pi\frac{|\bm{r}_{ij} + L \bm{\lambda}|^2}{L^2} \right)\left[ \frac{\cos (2k_{\rm F} |\bm{r}_{ij} + L \bm{\lambda}|)}{|\bm{r}_{ij} + L \bm{\lambda}|^3} - \frac{\sin (2k_{\rm F} |\bm{r}_{ij} + L \bm{\lambda}|)}{2k_{\rm F}|\bm{r}_{ij} + L \bm{\lambda}|^4} \right], \\ \nonumber
J_{ij}^{\rm long(3D)} &=& \frac{\pi^{\frac{3}{2}}}{\sqrt{3}cL^3}\sum_{\bm{h}} \frac{e^{2\pi i \frac{\bm{h}}{L}}\cdot \bm{r}_{ij}}{h} \Bigg\{ h_+E_1(\pi h_+^2) - h_-E_1(\pi h_-^2) \\
&-&\frac{1}{2k_{\rm F}L} \left[ \pi h_+^2 E_1 (\pi h_+^2) - \pi h_-^2 E_1 (\pi h_-^2) - (e^{-\pi h_+^2} - e^{-\pi h_-^2}) \right] \Bigg\},
\end{eqnarray}
\end{widetext}
where $\bm{h} = m_a \bm{k}_a + m_b \bm{k}_b + m_c \bm{k}_c$ with $\bm{k}_a = (1,-1/\sqrt{3}, 0),~\bm{k}_b = (0,2/\sqrt{3},0),~\bm{k}_c = (0,0,1/c)$, $h=|\bm{h}|$, $h_\pm = h \pm L k_{\rm F}/\pi$, and $E_1(x) = \int_x^\infty dt e^{-t}/t$ is the exponential integral function. The sum $\sum_{\bm{h}}$ runs over all integers $m_\mu = -\infty,...,0,...\infty$ ($\mu = a,b,c$).

Then, we take the 2D limit, i.e., $c \to \infty$. The Ewald potential can be written as
\begin{equation}
J_{ij}^{\rm Ewald(2D)} =  -\frac{J_0}{\Gamma(\frac{3}{2})} \left(J_{ij}^{\rm short(2D)} + J_{ij}^{\rm long(2D)} \right).
\end{equation}
The short-range contribution in 2D has the same form as that in 3D, 
\begin{equation}
J_{ij}^{\rm short(2D)} = J_{ij}^{\rm short(3D)},
\end{equation}
while $\bm{\lambda}$ is restricted to the $a$-$b$ plane. In case of the long-range contribution, the summation in the $c$-direction is replaced by the integral,%
\begin{equation}
\frac{1}{c}\sum_{\bm{h}} = \frac{1}{c}\sum_{\bm{h}_{ab}} \sum_{\bm{h}_c} \underset{c \to \infty}{\to} \sum_{\bm{h}_{ab}} \int_{-\infty}^{\infty} dh_c ,
\end{equation}
where $\bm{h}_{ab} = m_a \bm{k}_a + m_b \bm{k}_b$ and $\bm{h}_{c} = m_c \bm{k}_c$. The long-range term $J_{ij}^{\rm long}$ then becomes
\begin{eqnarray}
\nonumber
J_{ij}^{\rm long(2D)} &&= \frac{\pi^{\frac{3}{2}}}{\sqrt{3}L^3}\sum_{\bm{h}_{ab}}e^{2\pi i \frac{\bm{h}}{L}\cdot \bm{r}_{ij}}  \int_{-\infty}^{\infty} dh_c  \\ 
&& \times\frac{1}{h}\Bigg\{ h_+E_1(\pi h_+^2) - h_-E_1(\pi h_-^2) \nonumber \\
&&-\frac{1}{2k_{\rm F}L} \Big[ \pi h_+^2 E_1 (\pi h_+^2)- \pi h_-^2 E_1 (\pi h_-^2) \nonumber \\
&& - (e^{-\pi h_+^2} - e^{-\pi h_-^2}) \Big] \Bigg\}.
\label{Jlong}
\end{eqnarray}
The numerical evaluation of the right-hand side of Eq. (\ref{Jlong}) requires the evaluation of the double integral ($E_1$ in the integrand contains an integral). In fact, the $\int_{-\infty}^{\infty} dh_c (h_-/h) E_1(\pi h_-^2)$ term has an $x \log x$-type singularity around $h_- = 0$. To properly take account of this singularity in the numerical integration for $h_c$ by using Simpson's rule, we take a finer mesh of $10^{-5}$ around $h_- = 0$, while we take $10^{-4}$ in other intervals.

\section{Monte Carlo results of the temperature dependences of several physical quantities in single-$q$ phase}\label{low_field}

\begin{figure}[t]
\includegraphics[clip,width=85mm]{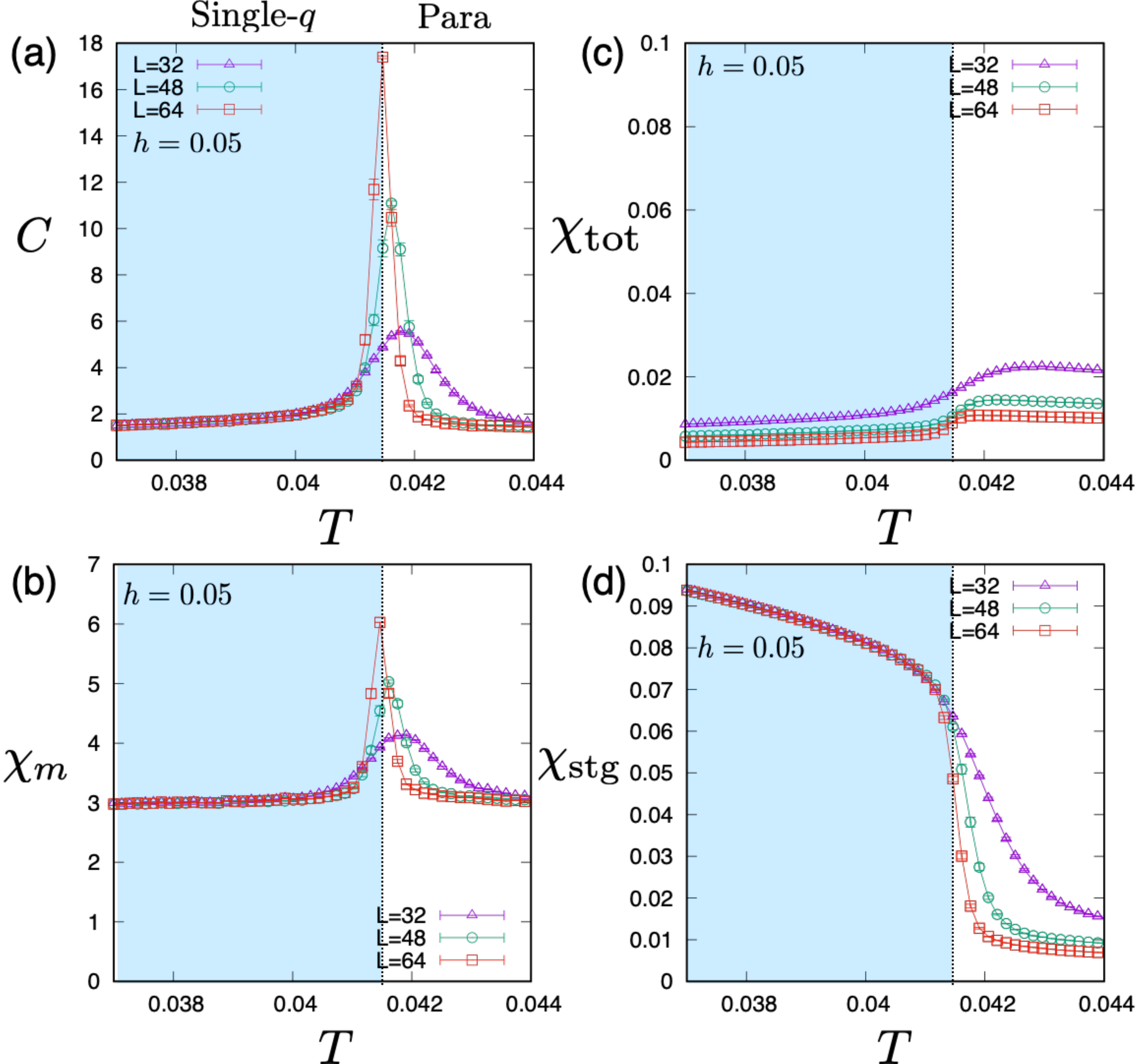}
\caption{
The temperature dependences of (a) the specific heat, (b) the uniform susceptibility, (c) the total scalar chirality, (d) the staggered scalar chirality at the low magnetic field of $h=0.05$, crossing in the phase diagram the paramagnetic and single-$q$ phases.
}
\label{tdep_low}
\end{figure}

In Figs. \ref{tdep_low} (a)-(d), we show the data at the low magnetic field of $h=0.05$ in the temperature range of $T\geq 0.037$. 
The specific heat $C$ shown in Fig. \ref{tdep_low} (a) exhibits a sharp peak at $T_{\rm c}^{\rm (low)}=0.0414$, corresponding to the phase transition between the paramagnetic and single-$q$ phases. 
The magnetic susceptibility $\chi_{m}$ shown in Fig. \ref{tdep_low} (b) exhibits a behavior more or less similar to that of the specific heat.
As can be seen from Fig. \ref{tdep_low} (d), the staggered scalar chirality $\chi_{\rm stg}$ is enhanced in the single-$q$ phase, while the total scalar chirality $\chi_{\rm tot}$ is not as shown in Fig. \ref{tdep_low} (c).
 
 \section{The $Z_3$-symmetry-breaking parameters}\label{rho3}
 
 \begin{figure*}[t]
\includegraphics[clip,width=170mm]{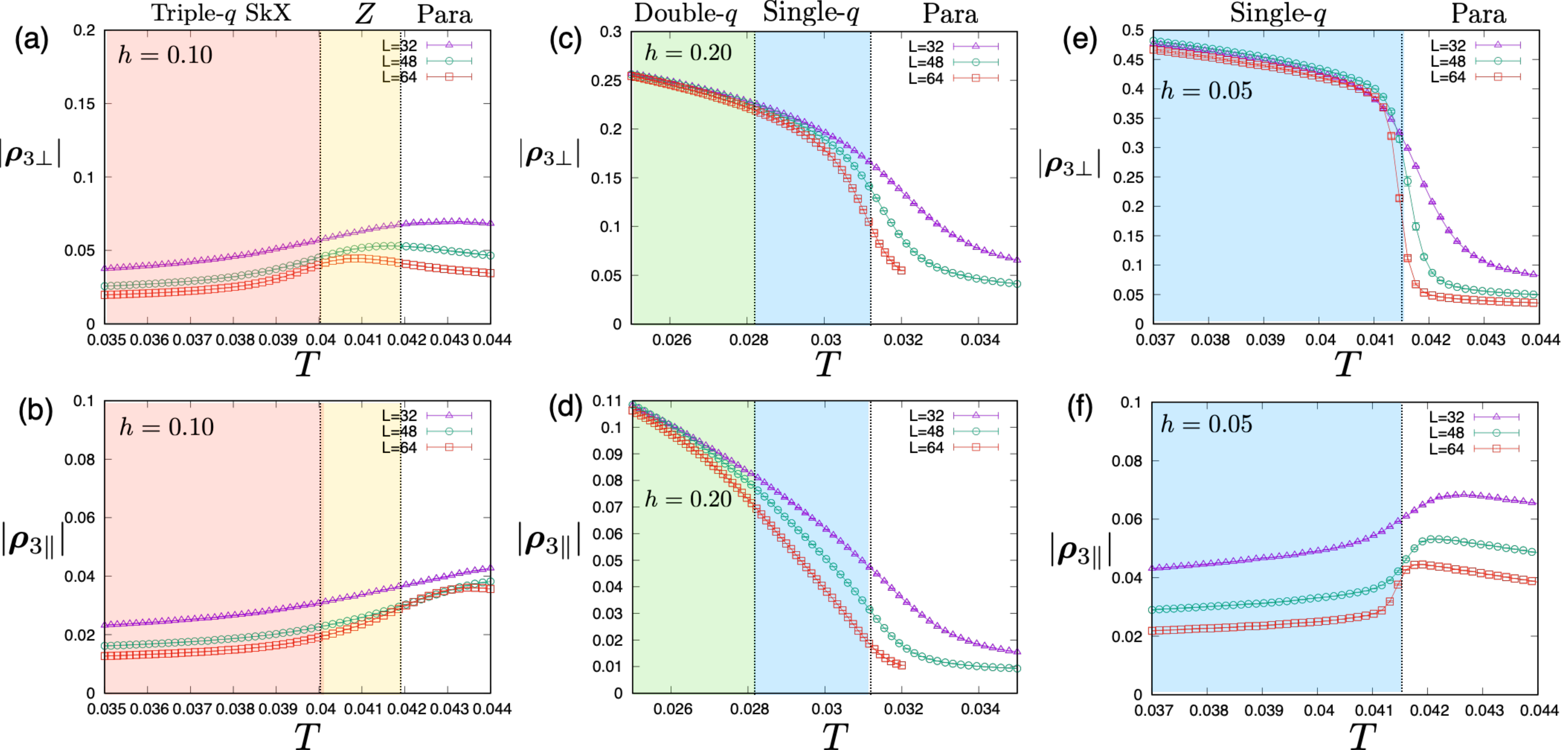}
\caption{
The temperature dependences of the absolute value of the perpendicular $Z_3$-breaking parameter $|\bm{\rho}_{3\perp}|$, and the absolute value of the parallel $Z_3$-breaking parameter $|\bm{\rho}_{3\parallel}|$. 
(a) $|\bm{\rho}_{3\perp}|$ and (b) $|\bm{\rho}_{3\parallel}|$ at the intermediate magnetic field of $h=0.10$, crossing in the phase diagram the paramagnetic, $Z$, and triple-$q$ SkX phases. 
(c) $|\bm{\rho}_{3\perp}|$ and (d) $|\bm{\rho}_{3\parallel}|$ at the high magnetic field of $h=0.20$, crossing in the phase diagram the paramagnetic, single-$q$, and double-$q$ phases.
(e) $|\bm{\rho}_{3\perp}|$ and (f) $|\bm{\rho}_{3\parallel}|$ at the low magnetic field of $h=0.05$, crossing in the phase diagram the paramagnetic and single-$q$ phases.
}
\label{tdep_3fold}
\end{figure*}

The $Z_3$-symmetry-breaking parameters for the spin $xy$-components and the $z$-component are defined by
\begin{eqnarray}
\bm{\rho}_{3\perp} &= s_\perp (\bm{q}_1^*) \bm{e}_1 +  s_\perp (\bm{q}_2^*) \bm{e}_2 + s_\perp (\bm{q}_3^*) \bm{e}_3, \label{rho3v} \\
\bm{\rho}_{3\parallel} &= s_\parallel (\bm{q}_1^*) \bm{e}_1 +  s_\parallel (\bm{q}_2^*) \bm{e}_2 + s_\parallel (\bm{q}_3^*) \bm{e}_3 \label{rho3p} ,
\end{eqnarray}
where $\bm{e}_1 = (1,0)$, $\bm{e}_2 = (-1/2,\sqrt{3}/2)$ and $\bm{e}_3 = (-1/2,-\sqrt{3}/2)$  are the unit vectors, $s_\perp (\bm{q})$ and $s_\parallel (\bm{q})$ are the instantaneous spin structure factors given by Eqs. (\ref{insta_perp}) and (\ref{insta_para}) without the thermal average. These order parameters $\bm{\rho}_{3\perp}$ and $\bm{\rho}_{3\parallel}$ are two-component vectors defined in the two-dimensional triangular order-parameter space and represent the extent of the $Z_3$-symmetry breaking. Hereafter, we call them the perpendicular and parallel $Z_3$-breaking parameters. $\bm{\rho}_{3\parallel}$ is regarded as the order parameter of the double-$q$ phase because the $Z_3$ symmetry is not broken for the spin $z$-component in the other phases.

Figures \ref{tdep_3fold} (a)-(f) show the $Z_3$-breaking parameters at three typical magnetic fields: i.e., the intermediate field of $h=0.10$ crossing in the phase diagram the paramagnetic, $Z$, and triple-$q$ SkX phases, the high field of $h=0.20$ crossing in the phase diagram the paramagnetic, single-$q$, and double-$q$ phases, and the low field of $h=0.05$ crossing in the phase diagram the paramagnetic and single-$q$ phases. The perpendicular $Z_3$-breaking parameter $|\bm{\rho}_{3\perp}|$ is enhanced in both the single-$q$ and double-$q$ phases, while the parallel $Z_3$-symmetry breaking parameter $|\bm{\rho}_{3\parallel}|$ is enhanced only in the double-$q$ phase.

\section{Distribution functions of the $Z_3$-breaking parameter}\label{histogram}

In this section, we show our MC results of the distributions of the $Z_3$-breaking parameter in various phases. 
Attention is paid to the issue of the existence/nonexistence of the RSB recently identified in the 3D RKKY system where the single-$q$ state macroscopically coexists with the triple-$q$ SkX state or the double-$q$ state \cite{mitsumoto2021replica}.

\begin{figure*}[t]
\includegraphics[clip,width=170mm]{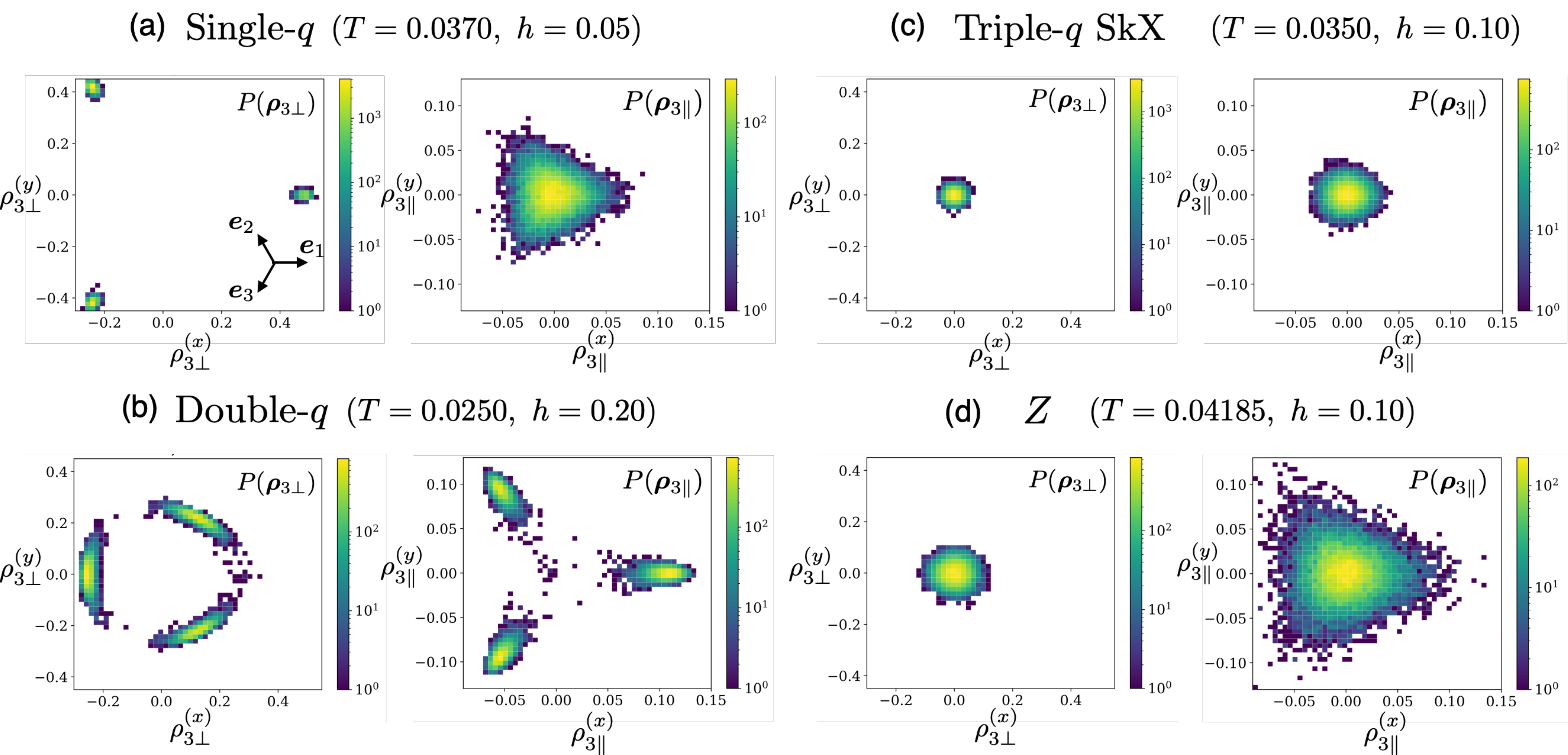}
\caption{
The histograms of the perpendicular and parallel $Z_3$-breaking parameters in (a) the single-$q$ phase, (b) the double-$q$ phase, (c) the triple-$q$ SkX phase, and (d) the $Z$ phase.
}
\label{srho_hist}
\end{figure*}

 The non-RSB feature can also be seen in the histograms of the perpendicular and parallel $Z_3$-breaking parameters, $P(\bm{\rho}_{3\perp})$ and $P(\bm{\rho}_{3\parallel})$, shown in Fig. \ref{srho_hist}. The single-$q$ state is characterized by the intensities in $P(\bm{\rho}_{3\perp})$ located at the corners of the triangular region, as shown in the left panel of Fig. \ref{srho_hist} (a). As can be seen from Figs. \ref{srho_hist} (b)-(d), no such intensity is detected in either the double-$q$, triple-$q$, or $Z$ phases.

\section{Derivation of the effective many-body interactions due to thermal fluctuations}\label{derivation}

 In this section, we rigorously derive from the microscopic Heisenberg spin Hamiltonian the effective many-body interactions due to thermal fluctuations by performing a kind of coarse graining procedure \cite{fisher1983scaling, kawamura1988renormalization}. We consider the $n$-component classical vector spins $\bm{S}_i = (S_i^1,S_i^2,...,S_i^n)~(i=1,2,...,N)$ normalized as $|\bm{S}_i| = 1$ on an arbitrary lattice in general $d$ dimensions, interacting via only the two-body interactions $J_{ij}$ whose energy scale is $J_0$. It includes the RKKY isotropic Heisenberg model treated in the main text as a special case.
The Hamiltonian under the site-dependent generalized magnetic field $\bm{h}_i$ is given by
\begin{equation}
\beta H = -\frac{1}{2}\sum_{i=1}^N\sum_{j \neq i} J_{ij} \bm{S}_i \cdot \bm{S}_j - \sum_{i=1}^N \bm{h}_i \cdot \bm{S}_i.
\end{equation}
Introducing the dimensionless matrix $ \tilde{J}_{ij} = P_0 \delta_{ij} + J_{ij}/J_0$, where $P_0>0$ is a positive constant sufficiently large to make $\tilde{J}_{ij}$ positive definite, one can apply the Hubbard-Stratonovich transformation,
\begin{widetext}
\begin{eqnarray}
\nonumber
Z &=& \underset{\{ \bm{S_i} \}}{\rm Tr} \exp \left( \frac{\beta J_{0}}{2} \sum_{i=1}^N\sum_{j =1}^N \tilde{J}_{ij} \bm{S}_i \cdot \bm{S}_j + \beta \sum_{i=1}^N \bm{h}_i \bm{S}_i \right) \\ \nonumber
&=& \frac{1}{A} \int \prod_{i=1}^N (d \bm{\phi}_i) \underset{\{ \bm{S_i} \}}{\rm Tr} \exp \left( -\frac{\beta J_{0}}{2} \sum_{i=1}^N\sum_{j =1}^N (\tilde{J}^{-1})_{ij} \bm{\phi}_i \cdot \bm{\phi}_j 
+ \sum_{i=1}^N(\beta J_0 \bm{\phi}_i + \beta \bm{h}_i) \cdot \bm{S}_i \right) \\
&=& \frac{1}{A} \int \prod_{i=1}^N (d \bm{\phi}_i) e^{-\beta H_{\rm eff}},
\end{eqnarray}
\end{widetext}
where $\beta=1/T$ is the inverse temperature, and $H_{\rm eff}$ is the effective Hamiltonian given by,
\begin{align}
\nonumber
H_{\rm eff} &= J_0 \Bigg[ \frac{1}{2} \sum_{i=1}^N \sum_{j=1}^N (\tilde{J}^{-1})_{ij} \bm{\phi}_i \cdot \bm{\phi}_j \\
&- \frac{1}{\beta J_0} \log \underset{\{ \bm{S_i} \}}{\rm Tr} \prod_{i=1}^N e^{(\beta J_0 \bm{\phi}_i + \beta \bm{h}_i) \cdot \bm{S}_i } \Bigg], \label{Heff1}
\end{align}
with $A = ((2\pi/\beta J_0)^N /|{\rm det} \tilde{J}^{-1}|)^{n/2}$.

 The auxiliary field $\bm{\phi}_i$ and the $n$-component spin $\bm{S}_i$ are related as,
\begin{equation}
\langle S_i^\mu \rangle = \frac{\partial \log Z}{\partial (\beta h_i^\mu)} = \sum_{j=1}^N (\tilde{J}^{-1})_{ij} \langle \phi_j^\mu \rangle,
\label{S-phi}
\end{equation}
which can easily be confirmed by integrating by parts after shifting the derivative with respect to $\beta h_i^\nu$ to the one with respect to $\beta J_0 \phi_i^\nu$. Since the field $\bm{\phi}_i$ is of unconstrained length, Eq. (\ref{S-phi}) suggests that $\bm{\phi}_i$ can be regarded as the coarse-grained spin variable.

We consider below the zero field case for simplicity. The trace in the second term of Eq. (\ref{Heff1}) can be taken as
\begin{equation}
\log \underset{\{ \bm{S_i} \}}{\rm Tr} \prod_{i=1}^N e^{\beta J_0 \bm{\phi}_i \cdot \bm{S}_i } = \sum_{i=1}^N \log \left( \frac{(2\pi)^{\frac{n}{2}} I_{\frac{n}{2}-1}(\beta J_0 \phi_i)}{(\beta J_0 \phi_i)^{\frac{n}{2}-1}} \right),
\end{equation}
where
\begin{equation}
I_{\nu}(x) = \sum_{m=0}^\infty \frac{1}{m ! \Gamma (m + \nu + 1)} \left( \frac{x}{2} \right)^{2m + \nu} ,
\end{equation}
is the modified Bessel function and $\phi_i$ is the absolute value of $\bm{\phi}_i$. We then expand the effective Hamiltonian up to the fourth order in $\bm{\phi}_i$ to get,
\begin{align}
\nonumber
H_{\rm eff} &= J_0 \Bigg[ \frac{1}{2} \sum_{i=1}^N \sum_{j=1}^N (\tilde{J}^{-1})_{ij} \bm{\phi}_i \cdot \bm{\phi}_j \\
&- \frac{\beta J_0}{2n} \sum_{i=1}^N \phi_i^2 + \frac{(\beta J_0)^3}{4n^2(n+2)} \sum_{i=1}^N \phi_i^4 \Bigg].
\end{align}
Using the translational symmetry, the effective Hamiltonian can be rewritten as
\begin{align}
\nonumber
H_{\rm eff} &= J_0 \Bigg[ \frac{1}{2} \sum_{\bm{q}}  \left( \tilde{J}^{-1}_{\bm{q}}  - \frac{\beta J_0}{n} \right) |\bm{\phi}_{\bm{q}}|^2 \\
&+ \frac{(\beta J_0)^3}{4n^2(n+2)} \sum_{\{\bm{q} \}}{}^\prime (\bm{\phi}_{\bm{q}_1} \cdot \bm{\phi}_{\bm{q}_2})(\bm{\phi}_{\bm{q}_3} \cdot \bm{\phi}_{\bm{q}_4}) \Bigg],
\end{align}
where $\bm{\phi}_{\bm{q}}$ and $\tilde{J}^{-1}_{\bm{q}}$ the Fourier transforms of $\bm{\phi}_i$ and $(\tilde{J}^{-1})_{ij}$,
\begin{eqnarray}
\bm{\phi}_{\bm{q}} &=& \frac{1}{N} \sum_{i=1}^N \bm{\phi}_i e^{-i \bm{q} \cdot \bm{r}_i},\\
\tilde{J}^{-1}_{\bm{q}} &=& \sum_{j=1}^N (\tilde{J}^{-1})_{ij} e^{-i \bm{q}\cdot \Delta \bm{r}_{ij}} ,
\end{eqnarray}
and $\sum_{\{\bm{q} \}}{}^\prime$ represents the restricted summation under the constraint $\bm{q}_1 + \bm{q}_2 + \bm{q}_3 + \bm{q}_4 = \bm{0}$. Note that the prefactor of the fourth-order term is positive. Indeed, the summation includes the $q$-space biquadratic interaction term with the positive coefficient which is of exactly the same form as the one used in Ref. \cite{hayami2017effective} if the relevant $\bm{q}$'s are restricted to the ordering wavevectors $\bm{q}^*$. Near the critical point $T_c$, the temperature is of order $J_0$, and $\beta J_0$ is of $O(1)$. Therefore, the many-body interactions due to thermal fluctuations have the same energy scale as the original two-body interaction, not containing the small parameter such as $(J_{sd}/\epsilon_{\rm F})$. Meanwhile, as the temperature is further lowered away from $T_c$, $\beta J_0$ becomes large, and the expansion of the effective Hamiltonian eventually breaks down.

\section{Evaluation of the small parameter $(J_{sd}/\epsilon_{\rm F})^2$ from the experimental data} \label{evaluation}

In this section, we numerically evaluate the typical value of the small parameter $(J_{sd}/\epsilon_{\rm F})^2$ from the available experimental data on Gd$_2$PdSi$_3$, a centrosymmetric metallic magnet exhibiting the SkX state in applied magnetic fields. For this purpose, we first evaluate the Fermi energy $\epsilon_F$ from the angle-resolved photoemission spectroscopy data and the energy scale of the RKKY interaction $J_0$ from the high-temperature susceptibility data. Then, $J_{sd}$ is related to $J_0$ and $\epsilon_F$ as \cite{yosida1957magnetic}
\begin{equation}
J_0 \approx 9\pi \epsilon_{\rm F} \left(\frac{J_{sd}}{\epsilon_{\rm F}}\right)^2,
\label{J0-Jsd}
\end{equation}
where the half-filling condition is assumed for simplicity.

 The Fermi energy $\epsilon_F$ of Gd$_2$PdSi$_3$ might be estimated from the angle-resolved photoemission spectroscopy measurements. The energy of the Fermi level measured from the bottom of the conduction band was reported to be $\epsilon_{\rm F} = 0.50$eV $= 5.8\times 10^{3}$K \cite{inosov2009electronic}.

 We estimate $J_0$ of Gd$_2$PdSi$_3$ by comparing the Curie-Weiss temperature determined experimentally from the high-temperature susceptibility measurements and the one determined theoretically from the high-temperature expansion of the 2D RKKY model. At high temperatures, the inverse magnetic susceptibility exhibits a linear dependence on the temperature $T$ as,
\begin{equation}
\chi^{-1} \propto (T - \Theta_{\rm p}) , \label{inv_mag}
\end{equation}
where $\Theta_p$ is the Curie-Weiss temperature. According to Ref. \cite{mallik1998magnetic}, $\Theta_p$ of Gd$_2$PdSi$_3$ is positive, i.e., ferromagnetic, and its value was estimated to be $\sim$19K. Approximating Gd$_2$PdSi$_3$ by the 2D RKKY model on the triangular lattice described by Eq.(1) of the main text, we derive the high-temperature form of the inverse susceptibility by means of the standard high-temperature expansion. Expansion to the first nontrivial order yields,

\begin{equation}
\chi^{-1} = \frac{3}{4} \left[ T - \frac{1}{3} \sum_{j=1}^N J_{ij} \right] .
\label{chi_theory}
\end{equation}
From Eqs. (\ref{inv_mag}) and (\ref{chi_theory}), we get
\begin{equation}
\Theta_{\rm p} (J_0, k_F) = \frac{1}{3} \sum_{j=1}^N J_{ij}(J_0, k_F).
\label{CurieWeiss}
\end{equation}
 Note that the RKKY interaction $J_{ij}$, and consequently $\Theta_p$, depend not only on $J_0$ but also on $k_F$. Generally speaking, the sign of $\Theta_p$ could be either positive (ferromagnetic) or negative (antiferromagnetic) depending on the $k_F$-value. In view of the experimental observation of the positive $\Theta_p$ for Gd$_2$PdSi$_3$, we consider here the $k_F$-region corresponding to the incommensurate spiral order with the positive $\Theta_p$ in the ground state. Referring to the ground-state phase diagram shown in Fig.1 of the main text, we choose here two candidate $k_F$ values, i.e., $k_{\rm F} = 2\pi/2.71$ and $2\pi/2.61$, the former being exactly the $k_F$-value employed in the MC simulation of the main text, and numerically estimate the right-hand side of Eq. (\ref{CurieWeiss}) based on the Ewald-sum method with $L=32$ (the $L$-dependence turns out to be almost negligible). From the relation $\Theta_p=19$K, we numerically estimate $J_{0}$ to be 250K and 59K, and by using Eq. (\ref{J0-Jsd}) the small parameter $(J_{sd}/\epsilon_{\rm F})^2$ to be $1.5\times 10^{-3}$ and $3.6\times 10^{-4}$ for $k_{\rm F} = 2\pi/2.71$ and $2\pi/2.61$, respectively. Hence, the parameter $(J_{sd}/\epsilon_{\rm F})^2$ is likely to be quite small in the metallic magnet Gd$_2$PdSi$_3$, around $10^{-3} - 10^{-4}$. This observation indicates that the fourth-body interaction arising from the higher-order perturbation of Ref. \cite{hayami2017effective} would be smaller than the quadratic RKKY interaction by the factor of $10^{-3} - 10^{-4}$ in typical weak-coupling metals.

\bibliography{rkky_2d_ref}

\end{document}